\documentclass[sigplan,10pt,screen,authorversion]{acmart}
\renewcommand\footnotetextcopyrightpermission[1]{} 

\usepackage{listings}
\usepackage{tcolorbox}
\usepackage{graphicx}
\usepackage{caption}
\usepackage{subcaption}
\usepackage{float}
\usepackage{tikz}
\definecolor{codegreen}{rgb}{0,0.6,0}
\definecolor{codegray}{rgb}{0.5,0.5,0.5}
\definecolor{codepurple}{rgb}{0.58,0,0.82}
\definecolor{lightgray}{rgb}{0.9,0.9,0.9}
\definecolor{darkblue}{RGB}{0, 0, 140}
\definecolor{DARKBLUE}{RGB}{0, 0, 140}

\lstdefinestyle{nanos_style}{
    basicstyle=\ttfamily\scriptsize,
    numberstyle=\tiny\color{codegray},
    breakatwhitespace=false,
    breaklines=true,
    captionpos=b,
    keepspaces=true,
    numbers=left,
    numbersep=2pt,
    showspaces=false,
    showstringspaces=false,
    showtabs=false,
    tabsize=2,
	language=C,
	float
}

\AtBeginDocument{%
  \providecommand\BibTeX{{%
    \normalfont B\kern-0.5em{\scshape i\kern-0.25em b}\kern-0.8em\TeX}}}

\setcopyright{none}
\copyrightyear{2022}
\acmYear{2022}
\acmDOI{TBD}

\newcommand\copyrighttext{%
  \footnotesize\sffamily Copyright \copyright 2022. This manuscript version is made available under the CC-BY-NC-ND 4.0 license.}
\newcommand\copyrightnotice{%
\begin{tikzpicture}[remember picture,overlay]
\node[anchor=south,yshift=50pt] at (current page.south) {{\parbox{\dimexpr\textwidth-\fboxsep-\fboxrule\relax}{\copyrighttext}}};
\end{tikzpicture}%
}

\begin{document}

\title[Accelerating Task-based Iterative Applications]{Accelerating Task-based Iterative Applications}

\author{David Álvarez}
\email{david.alvarez@bsc.es}
\orcid{0000-0002-4607-1627}
\affiliation{%
  \institution{Barcelona Supercomputing Center (BSC)}
  \city{Barcelona}
  \country{Spain}
}

\author{Vicenç Beltran}
\email{vbeltran@bsc.es}
\orcid{0000-0002-3580-9630}
\affiliation{%
  \institution{Barcelona Supercomputing Center (BSC)}
  \city{Barcelona}
  \country{Spain}
}

\renewcommand{\shortauthors}{David Álvarez and Vicenç Beltran}

\begin{abstract}
Task-based programming models have risen in popularity as an alternative to traditional fork-join parallelism.
They are better suited to write applications with irregular parallelism that can present load imbalance.
However, these programming models suffer from overheads related to task creation, scheduling and dependency management, limiting performance and scalability when tasks become too small.
At the same time, many HPC applications implement iterative methods or multi-step simulations that create the same directed acyclic graphs of tasks on each iteration.

By giving application programmers a way to express that a specific loop is creating the same task pattern on each iteration, we can create a single task DAG once and transform it into a cyclic graph.
This cyclic graph is then reused for successive iterations, minimizing task creation and dependency management overhead.
This paper presents the \textit{taskiter}, a new construct we propose for the OmpSs-2 and OpenMP programming models, allowing the use of directed cyclic task graphs (DCTG) to minimize runtime overheads.
Moreover, we present a simple \textit{immediate successor} locality-aware heuristic that minimizes task scheduling overhead by bypassing the runtime task scheduler.

We evaluate the implementation of the \textit{taskiter} and the \textit{immediate successor} heuristic in 8 iterative benchmarks.
Using small task granularities, we obtain an average speedup of 3.7x over the reference OmpSs-2 implementation and an average of 5x and 7.46x speedup over the LLVM and GCC OpenMP runtimes, respectively.
\end{abstract}

\ccsdesc[500]{Computing methodologies~Parallel programming languages}

\keywords{taskiter, tasks, runtimes, openmp, ompss-2}

\maketitle

\fancyhead[L]{\sffamily\footnotesize{}}
\fancyhead[R]{\sffamily\footnotesize{David Álvarez and Vicenç Beltran}}
\fancyfoot[C]{\vspace{1em}\sffamily\thepage}
\copyrightnotice

\section{Introduction}

Task-based programming models have become popular since they are better suited than fork-join models to uncover parallelism from dynamic and irregular applications.
Moreover, data-flow execution models, which use tasks and dependencies, can expose more parallelism in irregular applications, provide natural load balancing,
and leverage dependency information to improve data locality through smarter scheduling policies.
However, tasks suffer from creation, scheduling, and dependency management overheads, forcing programmers to balance task granularity.
We define \textit{task granularity} as the duration of each task in an application \cite{Granularity}.

When the granularity is too small, task creation, scheduling and dependency management become a bottleneck \cite{OpenMPGranularities, GranularityAnalysis2019, GranularityAnalysis2015, NavarroGranularity}, and tasks cannot be created fast enough to feed all cores.
This situation produces two adverse effects which hinder performance:
First, some cores remain idle, as not enough work is being created.
Second, as the number of tasks ready to execute is very low, there is little chance to apply locality-aware scheduling policies.
However, when tasks are too coarse, there may not be enough work to feed all cores, the program can suffer from load imbalance,
and locality-aware scheduling policies may lose effectiveness as working sets grow.

Thus, we want to create tasks in a \textit{balanced} region, where granularity is not too fine nor too coarse.
This is normally achieved through granularity tuning, but there are relevant situations where tuning is impossible.
For example, when the problem size is too small or when scaling-out an application. In these cases, it is critical that the runtime efficiently supports small task granularities.

Task-based programming models have been optimized over time to minimize these task management overheads~\cite{advancedsynchronizationtechniques, TurboBLYSK, tbb-oh}.
There are three main overhead sources in a task-based runtime: task creation, scheduling and dependency management.
Task creation is generally optimized using scalable memory allocators~\cite{jemalloc, hoard}.
Task scheduling is optimized with scalable scheduling techniques, such as work-stealing~\cite{Yang2018} or delegation-based schedulers~\cite{advancedsynchronizationtechniques}.
Finally, task dependency management requires fine-grained locking or wait-free implementations to achieve good performance.
However, these optimizations may not be enough to achieve competitive performance when very fine-grained tasks are needed.

At the same time, many HPC applications present an iterative pattern, creating the same tasks with the same dependencies for each iteration.
This results in identical tasks and dependency graphs that are concatenated one after the other.
For example, this happens in iterative methods and solvers, machine learning training phases and multi-step simulations.
As such, iterative programs can spend a significant amount of time creating, scheduling and managing tasks and dependencies that are the same for each iteration.

This paper presents and implements two techniques that drastically reduce the main sources of runtime overhead in task-based applications.
First, we propose a new \textit{taskiter} construct for the OmpSs-2~\cite{bsc2020ompss2} and OpenMP~\cite{openmp51} programming models.
The \textit{taskiter} construct annotates loops where each iteration generates the same directed acyclic graph (DAG) of tasks
and dependencies.
The runtime system then leverages this information to construct a directed cyclic task graph (DCTG) based on the DAG of the first iteration.
Dependencies between different iterations are considered and linked in this new directed cyclic graph.
In the DCTG, task descriptors and dependency structures are reused for each iteration, drastically reducing task creation and dependency
management overheads for any iterations after the first one.
It is worth noting that the \textit{taskiter} construct does not create any implicit barriers between iterations or after the construct, allowing it to be transparently mixed with successor or predecessor tasks or \textit{taskiter} constructs.

Secondly, we present a new \textit{immediate successor} scheduling technique that preserves data locality while drastically reducing scheduling overheads by bypassing the scheduler.
Unlike the \textit{taskiter}, this technique is not restricted to iterative applications.

Finally, we will show in the evaluation how both contributions present a particular synergy which results in significant performance improvements
for small granularities.

Specifically, our contributions are as follows:
\begin{enumerate}
	\item We propose a new construct for the OmpSs-2 programming model, and show how it could be adapted to fit the OpenMP standard.
	\item We propose a new scheduling policy designed to forego most of the scheduling overhead and maximize data locality.
	\item We implement both the taskiter construct and the scheduling policy on the reference implementation of OmpSs-2.
	\item We evaluate the taskiter construct on 8 iterative benchmarks and find an average speedup of 3.7x for small task granularities.
\end{enumerate}

The rest of this document is structured as follows:
Section \ref{sec:related} reviews the current state of the art,
Section \ref{sec:taskiter} introduces the \textit{taskiter} construct
and Section \ref{sec:immediate} introduces the \textit{immediate successor} technique.
Then, we evaluate our contributions in Section \ref{sec:evaluation}, and present the conclusions in Section \ref{sec:conclusions}.

\section{Related Work}
\label{sec:related}

The effects of task granularity on application performance have been thoroughly studied in literature
\cite{OpenMPGranularities, GranularityAnalysis2019, GranularityAnalysis2015, NavarroGranularity}.
Moreover, several proposals to reduce task management and scheduling overhead have been proposed.

\subsection{Task management overhead}
\label{sec:related-task}

Some works have focused on optimizations that can be applied to task-based runtimes
to reduce synchronization overheads and scale better \cite{advancedsynchronizationtechniques, tbb-oh}.
These techniques complement model-oriented solutions as the taskiter.

Other approaches have focused in reducing task overheads by decreasing the total number of tasks that have to be created.
Worksharing tasks \cite{WorksharingMarcos} and Chapel's \texttt{coforall} construct~\cite{chapel-coforall} can parallelize all iterations
from a loop using a single task, reducing their overhead. Similarly, Index Launches~\cite{indexlaunch} can automatically compact several task
launches in a loop without need for explicit annotation. Polytasks~\cite{polytasks} also merge several similar tasks when they are
created at the same time, provided tasks are managed through queues.
These approaches reduce the total number of tasks created by an application. In contrast, the proposed taskiter focuses on task reuse,
and both approaches can be freely combined, as they are complementary.

In \cite{TurboBLYSK}, the authors propose the \texttt{dep\_pattern} clause to cache data dependency patterns
reducing dependency management overhead. Our proposal goes further, not only caching dependency structures
but preventing task creation altogether. Moreover, the \texttt{dep\_pattern} clause must be placed on
a parent task, which in OpenMP would prevent placing dependencies between iterations to overlap their execution.

Another approach is task DAG caching, provided by the CUDA Graph API \cite{cuda}, which allows GPU programmers
to record a graph of kernel invocations and memory copy operations and re-invoke them, removing
a significant amount of overhead. The graph API was also motivated by applications with an iterative structure,
like machine learning training. However, CUDA Graphs require a barrier between iterations, which prevents the overlap of kernels from multiple iterations
and limits the applicability of policies like the immediate successor.
TBB Graphs~\cite{onetbb} also allow task graphs that contain cycles, but the programmer must explicitly instantiate all nodes and edges of a task graph
manually.

A task DAG caching proposal for OpenMP is the \texttt{taskgraph} clause for the \texttt{target} and \texttt{task} constructs~\cite{taskgraph}.
Similar to CUDA Graphs, authors present a way to record and re-play task DAGs for OpenMP tasks.
However, the approach requires task DAGs to be defined inside their own dependency domain with an implicit barrier at the end. Hence, dependencies
between tasks inside the construct and tasks outside it or in other re-plays are not allowed, breaking the \textit{data-flow} execution model.
The \texttt{taskgraph} model is a caching strategy and not a task DAG transformation like the one proposed in this paper.

These task caching proposals have the potential to improve the performance of iterative applications. However, as we will show in the experimental evaluation,
the taskiter construct can clearly outperform task caching approaches.

\subsection{Scheduling}

Many works have tackled overhead reduction in task scheduling, either through work-stealing \cite{Yang2018} or
global-queue techniques \cite{advancedsynchronizationtechniques}.
Moreover, there have been proposals for locality-aware scheduling, mainly focused on preventing remote accesses on NUMA systems \cite{locality-sched}.
However, our focus is not on optimal locality but on improving locality while simultaneously eliminating most of the scheduling overhead.

The philosophy for our scheduling work is similar to Cilk's work-first principle~\cite{cilk5}, which is to remove scheduling overheads from worker threads.
However, the heuristic we propose in Section \ref{sec:immediate} is adapted for data-flow applications, works well under any amount of available parallelsim,
and provides additional locality improvements.

\section{The \texttt{taskiter} construct}
\label{sec:taskiter}

\begin{figure}
	\centering
	\includegraphics[width=\columnwidth]{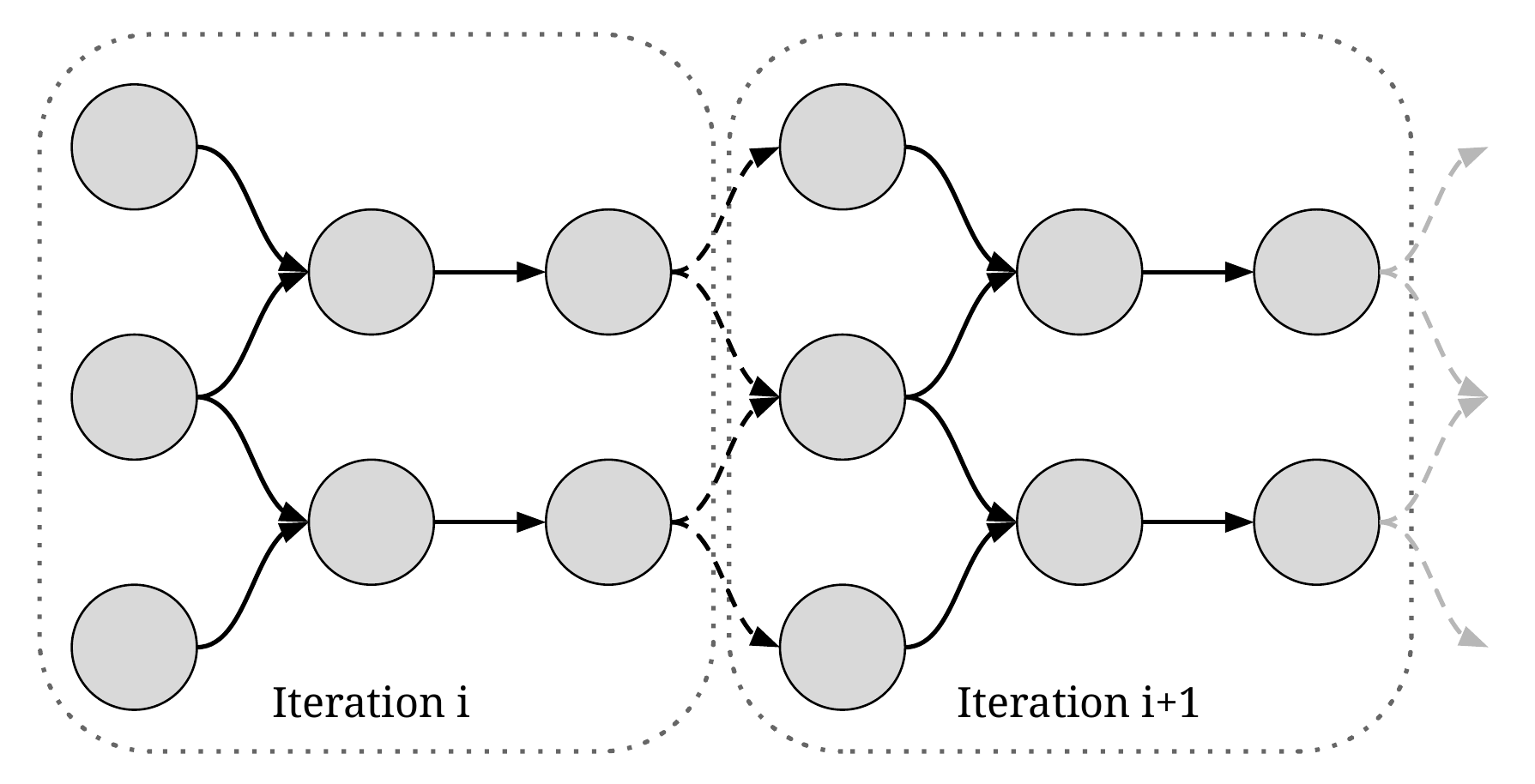}
	\caption{Iterative application}
	\label{fig:taskgraph-pre}
\end{figure}

Iterative applications often generate the same dependency graph on each iteration.
Dependencies always form a directed acyclic graph of tasks, enforcing restrictions on execution order
to maintain serial consistency.
Additionally, some task instances from an iteration $i$ will depend on task instances from previous iterations, as we avoid using global barriers.
An example of this pattern is shown in Figure \ref{fig:taskgraph-pre}, showing two iterations of an hypothetical iterative application.
Dependencies between task instances of the same iteration are shown in solid lines, and dashed lines indicate dependencies between iterations.

\begin{figure}
	\centering
	\includegraphics[width=.7\columnwidth]{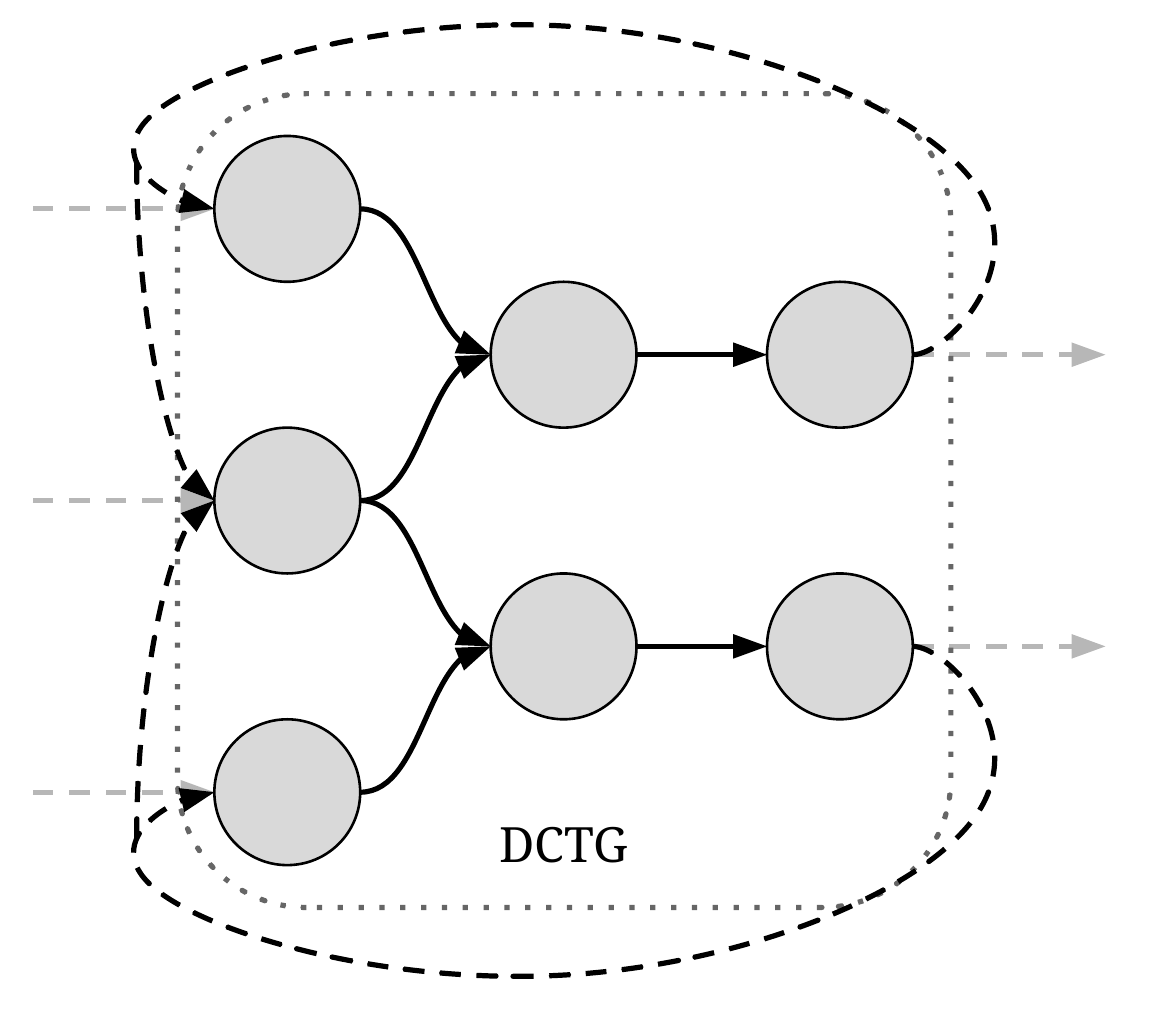}
	\caption{Iterative application after transformation to DCTG}
	\label{fig:taskgraph-post}
\end{figure}

The taskiter construct is designed to prevent creating and executing the same DAG for each iteration.
Instead, the programmer can express that a loop generates the same DAG $N$ times.
The programming model runtime will instead generate a directed \textbf{cyclic} task graph (DCTG), as shown in Figure \ref{fig:taskgraph-post}.
To build the DCTG the runtime executes the first iteration of the loop and generates a regular task DAG.
When the first iteration ends, the left and right sides of the DAG are connected, as shown in Figure \ref{fig:taskgraph-post}.
This representation is then used to execute the remaining $N-1$ iterations, minimizing task creation and dependency management overheads.

Moreover, the proposed DCTG representation is much more compact in memory than creating the task instances for every iteration.
This leads to a lower memory usage, which may otherwise be a problem for task-based programming models when the number of
task instances is very large.

This model makes it possible to execute task instances from different iterations simultaneously,
as the DCTG does not have any implicit barrier between iterations.
Additionally, dependencies from task instances on the first and last iterations can be matched to tasks outside the taskiter construct, maintaining the data-flow model.
Note that the task instances of the first iteration can be executed while building the DCTG, not introducing any performance penalty.

The syntax of the taskiter construct for OmpSs-2 is defined as the following:
\begin{tcolorbox}[width=\columnwidth, colback={lightgray}, arc=0mm, top=1mm, left=1mm, bottom=1mm, right=1mm]
\footnotesize{\texttt{\textbf{\#pragma oss taskiter}} \texttt{[clause [...]] new-line}}
\\\footnotesize{\hphantom{8mm}\texttt{loop}}
\end{tcolorbox}

The \textit{loop} can be any loop statement, provided it fulfills the following conditions:
\begin{enumerate}
	\item The dependency graph generated by the tasks inside the construct must remain constant for each iteration.
	However, nested tasks do not have this restriction.
	\item The program must remain valid if the code inside the loop body but outside any task is executed only once.
	This condition can be ignored if the \texttt{update} clause is specified, which we explain later on.
\end{enumerate}

The first condition is what the user is actually annotating with the taskiter construct: that the dependency graph for the loop repeats itself
and thus can be optimized to a cyclic graph. However, this only needs to be true for first-level tasks (created directly in the loop body), but not
for tasks created in deeper nesting levels, allowing irregularity between iterations.

The second condition allows the implementation to execute the loop body only once, and programs can generally be adapted to fulfill this condition by taskifying any code that is inside the loop body.

For example, we can apply the taskiter construct to an example Gauss-Seidel solver, which iterates through all blocks of a matrix in a wave-front pattern. This results in the code displayed in Listing \ref{lst:heat-tg}.
This code would fulfill the requirements of the taskiter, and it only requires the addition of Line 1 from the plain tasks version of this solver.

\begin{lstlisting}[language=C, label=lst:heat-tg, caption=Taskiter applied to a sample Gauss-Seidel Heat Equation application, float]
	#pragma oss taskiter
	for (int timestep = 0; timestep < N; ++timestep) {
		for (int R = 1; R < numRowBlocks - 1; ++R) {
			for (int C = 1; C < numColumnBlocks - 1; ++C) {
				#pragma oss task label(block computation) \
						in(reps[R-1][C]) in(reps[R+1][C]) \
						in(reps[R][C-1]) in(reps[R][C+1]) \
						inout(reps[R][C])
				computeBlock(rows, cols, R, C, matrix);
			}
		}
	}
\end{lstlisting}



Two new clauses can be combined with the proposed construct in OmpSs-2:
\begin{itemize}
	\item The \textbf{\texttt{unroll(n)}} clause performs loop unrolling, executing the initial \texttt{n} iterations instead of one.
	This clause can be used for loops with a regular dependency graph each \texttt{n} iterations. For example, a loop that behaves differently
	for even and odd iterations can be unrolled two times to generate the cyclic dependency graph. Moreover, with the unroll clause it is possible
	to have inter-iteration dependencies of distance up to \texttt{n}.
	\item A taskiter with the \textbf{\texttt{update}} clause will generate a cyclic dependency graph for its tasks only once, but the loop
	body will be executed for each iteration. Each time the loop body is executed, the parameters used to create each task instance will be recorded,
	allowing tasks in the generated DCTG to have different parameters for each iteration.
\end{itemize}

Use of the \texttt{taskloop} construct inside a taskiter is allowed, including taskloops with dependencies~\cite{taskloopdeps}.

The taskiter construct itself can also have dependencies, which can be used to express a dependency from the last iteration
of the taskiter to its sibling tasks.
This is demonstrated in Listing \ref{lst:tg-deps}, where using dependencies is convenient because the task in Line 13 can be created before the full DAG
of the taskiter in Line 5 is registered.

\begin{lstlisting}[language=C, label=lst:tg-deps, caption=Using dependencies between sibling tasks and tasks inside a taskiter region, float]
	int A;
	#pragma oss task out(A)
	A = 1;

	#pragma oss taskiter in(A) out(A)
	for (int i = 0; i < N; ++i) {
		#pragma oss task in(A)
		...
		#pragma oss task out(A)
		...
	}

	#pragma oss task in(A)
	print(A);
\end{lstlisting}

Task reductions are also supported inside a taskiter region because they do not require an enclosing \texttt{taskgroup} in OmpSs-2.

The loop that is transformed by the taskiter does not need to perform a constant number of iterations, and thus executing the next iteration
can depend on an arbitrary condition.
However, when the loop does not have a constant number of iterations, the programming model must guarantee that the condition is checked between
iterations and the taskiter is stopped when the condition becomes false.
Otherwise, when the number of iterations is a run-time constant, the programming model is free to overlap execution of tasks instances from as many
different iterations as the dependencies permit.

\subsection{Implementation}
When an OmpSs-2 compiler encounters a taskiter construct, it encapsulates one iteration of the following for loop
as a task.
That task is instantiated and passed to the OmpSs-2 runtime with the number of iterations to execute and a flag
indicating it is a taskiter.
This special task is queued for execution and will execute the loop's body once, creating any child tasks
and registering the initial DAG.
However, every child task instance will inherit an iteration counter from the taskiter to track how many times the task instance has to be executed.

\begin{figure}[h]
	\centering
	\includegraphics[width=1\columnwidth]{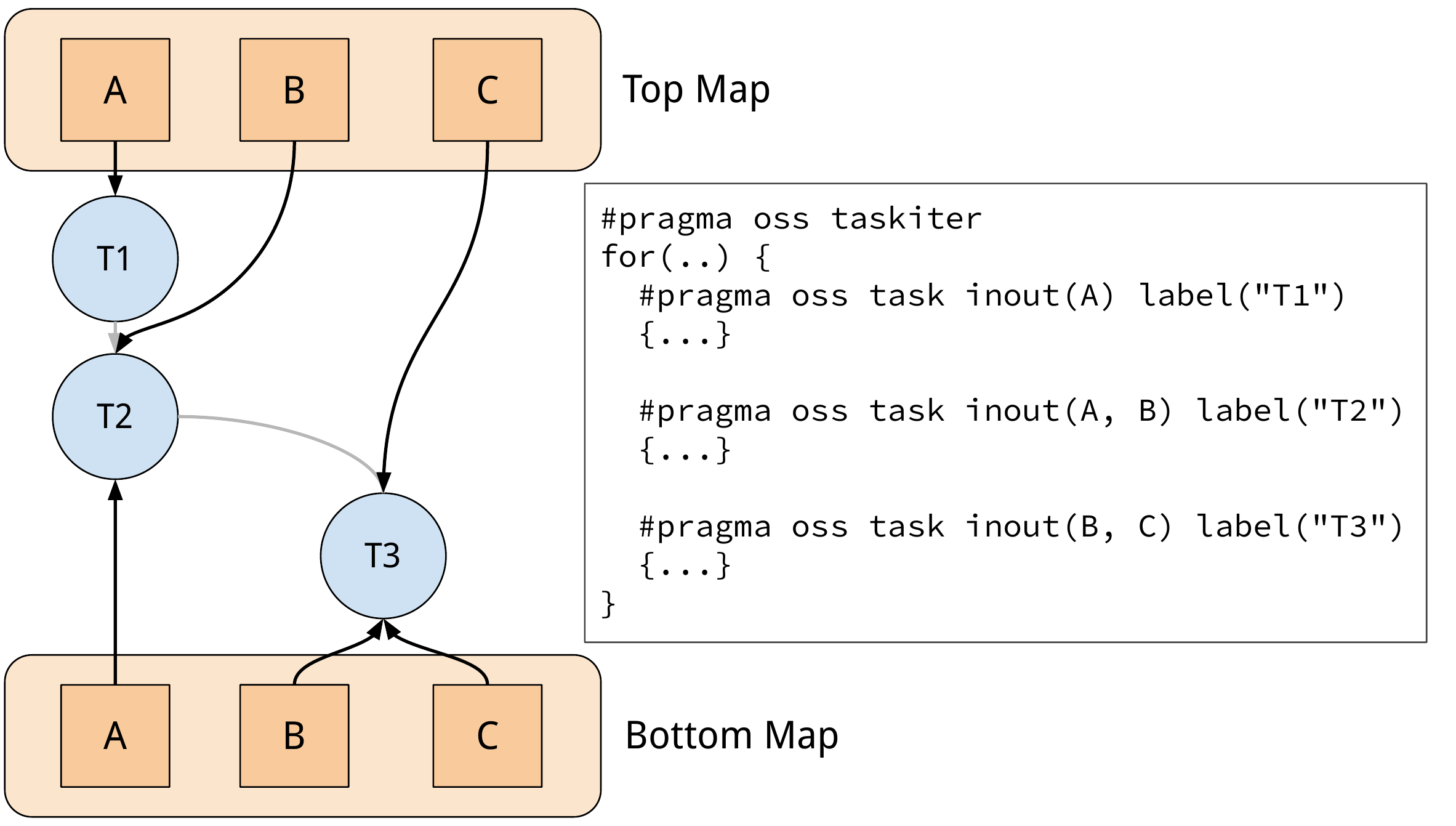}
	\caption{Implementation of the taskiter with top and bottom maps. Squares represent entries in the maps, and point to the first/last task to declare a dependency on a specific address.}
	\label{fig:impl}
\end{figure}

When the runtime has finished executing the first body of the loop, it will access the \textit{bottom map},
which is a data structure containing the last task that has declared a dependency on each memory location.
It will match those tasks to the \textit{top map}, which contains the first task that depends
on each memory location.
If locations match, there is a dependency from one iteration to the next, and we create an edge between
the last and first tasks depending on that location.
This edge is marked as crossing the iteration boundary.

An example of the top and bottom map structures is shown in Figure \ref{fig:impl}.
The pictured dependency graph corresponds to the attached code fragment.
In this example, for memory location \texttt{A}, the top map points to task instance \texttt{T1}, which is the first to declare a dependency on \texttt{A}.
Likewise, \texttt{T2} is the last task instance to declare a dependency on \texttt{A}.
Therefore, there is a cross-iteration dependency where \texttt{T1} depends on the previous iteration's \texttt{T2}.
With these data structures, finding the cross-iteration dependencies is reduced to matching all entries from the bottom map to the ones on the top map.

Whenever a child task finishes, it decreases its iteration counter, and unless it reaches zero, it will try to execute again
if its dependencies are satisfied.
Each task instance has two data structures that track outstanding dependencies: for even and odd iterations.
This way, we do not have to reinitialize the data structures after each iteration. We can track dependencies simultaneously for the current
and next iterations without inserting implicit barriers.
Moreover, this technique allows us to maintain the wait-freedom of the OmpSs-2 dependency system.

\subsection{Taskiter in OpenMP}
As OpenMP is the de-facto industry standard for shared-memory parallelism, we want to remark that the taskiter construct could be included
in the OpenMP standard with similar semantics to the OmpSs-2 version.
OpenMP's taskiter syntax would be analogous:
\begin{tcolorbox}[width=\columnwidth, colback={lightgray}, arc=0mm, top=1mm, left=1mm, bottom=1mm, right=1mm]
\footnotesize{\texttt{\textbf{\#pragma omp taskiter}} \texttt{[clause [...]] new-line}}
\\\footnotesize{\hphantom{8mm}\texttt{loop}}
\end{tcolorbox}

However, OpenMP does not allow dependencies across nesting levels.
In order to maintain the ability to express dependencies between the first and last iterations of the taskiter
and its sibling tasks, the semantics would be as follows:

\begin{itemize}
	\item By default, dependencies are scoped only to the taskiter, and the taskiter creates an implicit \texttt{taskgroup} region enclosing all iterations.
	\item If the \texttt{nowait} clause is specified, no implicit \texttt{taskgroup} region is created, and thus no implicit barrier. Note that because the \texttt{taskiter}
	is still an implicit task the execution of the loop body is deferred, and the encountering thread can progress.
	\item If the \texttt{inline} clause is specified, tasks are instantiated as if they were in line with the rest of the code, scoping dependencies to the enclosing region
	and allowing dependencies between the first and last iterations and its sibling tasks. In this case, the thread that encounters the \texttt{taskiter} must immediately execute the loop body.
\end{itemize}

Note we have to consider task reductions too, as they require an enclosing \texttt{taskgroup} in OpenMP.
Using a \texttt{taskgroup} inside a taskiter is not allowed, as including barriers defeats the taskiter's purpose.
Instead, the \texttt{task\_reduction} clause could be specified in a taskiter construct, and the corresponding \texttt{in\_reduction} clause
could be used in the tasks participating in the reduction.

To implement the taskiter construct in OpenMP, a similar strategy to the OmpSs-2 implementation can be taken.
For example, its implementation in LLVM OpenMP could re-use the transformation done by the OmpSs-2 LLVM compiler.
Additionally, both the GCC and LLVM implementations of OpenMP already maintain a bottom map structure, in the form
of a map of dependency chains, and a top map could be added in a similar fashion. This way, inter-iteration dependencies
can be calculated, and task execution can be controlled with iteration counters as in OmpSs-2.

\section{Immediate Successor}
\label{sec:immediate}

Using the taskiter, we can minimize the overhead of task creation and dependency management.
However, reducing those overheads shifts the contention to the remaining source of overhead: task scheduling.
After introducing the taskiter in our benchmarks, we observed that the scheduler could become the bottleneck, limiting application performance.
Specifically, the speed at which tasks are inserted and requested from the scheduler grows significantly, and so does contention on
the locking system of the scheduler.
OmpSs-2 currently features a delegation-based centralized scheduler~\cite{advancedsynchronizationtechniques}, but the same
contention can be observed in work-stealing implementations when there are few creators, which is common in data-flow applications.

This section presents a scheduling policy that maximizes data locality for task-based applications and can be applied
without acquiring any scheduler lock.
This way, we minimize the number of times any thread has to access the scheduler, reducing contention.

This heuristic is based on a straightforward \textit{successor locality} principle.
When one task has a dependency relation with another task, defined by the list of memory locations in their dependency clauses,
they probably share a part of their working set.
The reasoning behind this principle is straightforward. Data dependencies specify which memory locations
a task will access. If two tasks declare a dependency on the same location, both tasks will contain a memory reference to the same location,
sharing a part of their working set.

Formally, we define the working set of a task $t_i \in T$ as $W(t_i)$, representing the set of all memory locations that $t_i$ accesses during its execution.
Then, we can define a dependency relation, on which a task $t_1$ depends on a task $t_0$ as $t_1 \succ t_0$.
This denotes constraints in execution order and means that $t_1$ and $t_0$ share at least one memory location on the declared data dependencies.

Then, we propose the \textit{successor locality} principle:
\[ \forall t_0, t_1 \in T, t_1 \succ t_0 \to W(t_0) \cap W(t_1) \neq \emptyset \]

While it is possible to create a program on which the above statement is not valid,
it matches the patterns observable on most HPC applications written using a data-flow model.

Data locality is paramount when scheduling tasks because it allows
applications to exploit the memory hierarchy when the working sets fit in any cache level.

We can leverage this \textit{successor locality} principle to bypass the task scheduler while simultaneously preserving data locality.
We do this through the \textit{immediate successor} mechanism, which works as follows:

\begin{enumerate}
	\item Whenever a task finishes its execution, the worker thread executing it releases its dependencies and can mark one or more \textit{successor} tasks as ready.
	\item \label{step:markis} The first task with the highest priority marked as ready is kept into a local per-worker variable, becoming the \textit{immediate successor}.
	\item The rest of ready tasks (if any) are placed onto the scheduler for other workers to grab.
	\item If the worker has an \textit{immediate successor} task, the scheduler is bypassed, and the task is executed next.
\end{enumerate}

Note that we choose the first ready task amongst the ones with the highest priority as the immediate successor.
However, choosing the first one is arbitrary, as every ready task follows the successor locality principle.

While this policy is simple, it minimizes the number of times the scheduler is invoked, preventing contention.
Moreover, as we show during experimentation, it achieves significant speedups for some applications thanks to its locality-preserving property.

There is a trade-off when applying the immediate successor mechanism.
Bypassing the scheduler can be problematic when executing applications that rely on specific scheduling policies (for example, task priorities).
We can solve this issue by modifying step~\ref{step:markis} of the immediate successor algorithm, and adding a tunable probability that ready tasks
are not marked as immediate successors, allowing threads to enter the scheduler every once in a while.


\section{Experimental Evaluation}
\label{sec:evaluation}

To evaluate both the taskiter and the immediate successor (IS) policy,
we implemented both features on top of the reference implementation of the OmpSs-2 programming model.
Most changes are in the Nanos6~\cite{bsc2019nanos6} runtime, although we added compiler
support in the Mercurium~\cite{mercurium,mercurium2} source-to-source compiler and the LLVM-based OmpSs-2 compiler.
Our changes in the Nanos6 runtime only add a small constant overhead to dependency management for tasks inside a
taskiter, which does not depend on the number of iterations.

Software artifacts including the changes to the OmpSs-2 implementation, benchmark sources
and scripts to reproduce the experimental evaluation will be publicly available upon acceptance.

\subsection{Methodology}

The evaluation for the taskiter was conducted on a node equipped with an AMD EPYC 7742 (Rome) processor with 64 cores clocked
at 2.25 GHz and SMT disabled. The system has 1TiB of main memory at 3200MHz.
The software stack was composed of a CentOS Linux 8.1 distribution with a Linux 4.17 kernel.
The benchmarks were built with the Mercurium compiler and GCC 10.2.0.

The modified Nanos6 runtime was based on the publicly available OmpSs-2 2021.06 release, specifically on commit \texttt{39d5a111}.
We measured the performance when varying task granularity on a set of OmpSs-2 benchmarks:

\begin{itemize}
	\item A \textbf{Multisaxpy} benchmark, that performs a loop of $n$ Single $A\cdot X + Y$ kernels over two arrays.
	\item The \textbf{HPCG}~\cite{hpcg} (High Performance Conjugate Gradients) benchmark, with a fixed iteration count.
	\item The \textbf{HPCG (while)} variant, which is the HPCG benchmark with a variable iteration count, checking for convergence on each iteration.
	\item The \textbf{Heat} Gauss-Seidel equation solver that was showcased in Listing~\ref{lst:heat-tg}.
	\item The \textbf{Heat (while)}, which is the Heat with a variable iteration count, checking the residual on each iteration.
	\item A \textbf{N-Body} simulation that performs several timesteps of the interaction of forces in a particle system.
	\item A \textbf{Full-Waveform Inversion} proxy application used in exploration geophysics.
	\item The \textbf{HPCCG} proxy application that performs some of the relevant kernels from the HPCG benchmark without the pre-conditioning steps.
\end{itemize}

We run two experiments to evaluate the proposed extensions.
In the first experiment, we evaluate the performance of our taskiter and immediate successor policy using two task granularities:
one where tasks are small, simulating a strong scaling scenario, and another where granularity is optimal.
We evaluate each extension in isolation and combined with the rest of extensions.

In the second experiment, we do a granularity study for each benchmark comparing OmpSs-2 and OpenMP.

Finally, after presenting the results, we do a detailed analysis of the HPCG benchmark using execution traces.

\subsection{Experiment 1: Evaluation of proposed extensions}

\begin{figure*}
	\centering
	\begin{subfigure}[b]{1\textwidth}
		\centering
		\includegraphics[width=1\textwidth]{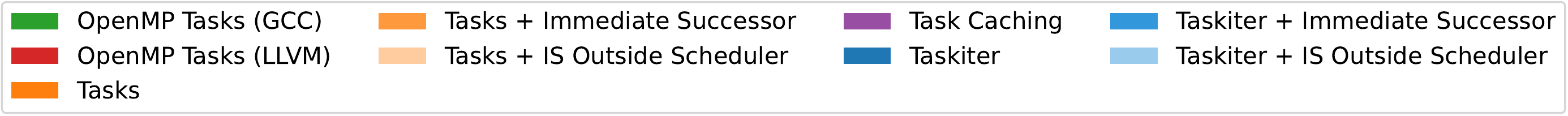}
	\end{subfigure}
	\begin{subfigure}[b]{.48\textwidth}
		\centering
		\includegraphics[width=\columnwidth]{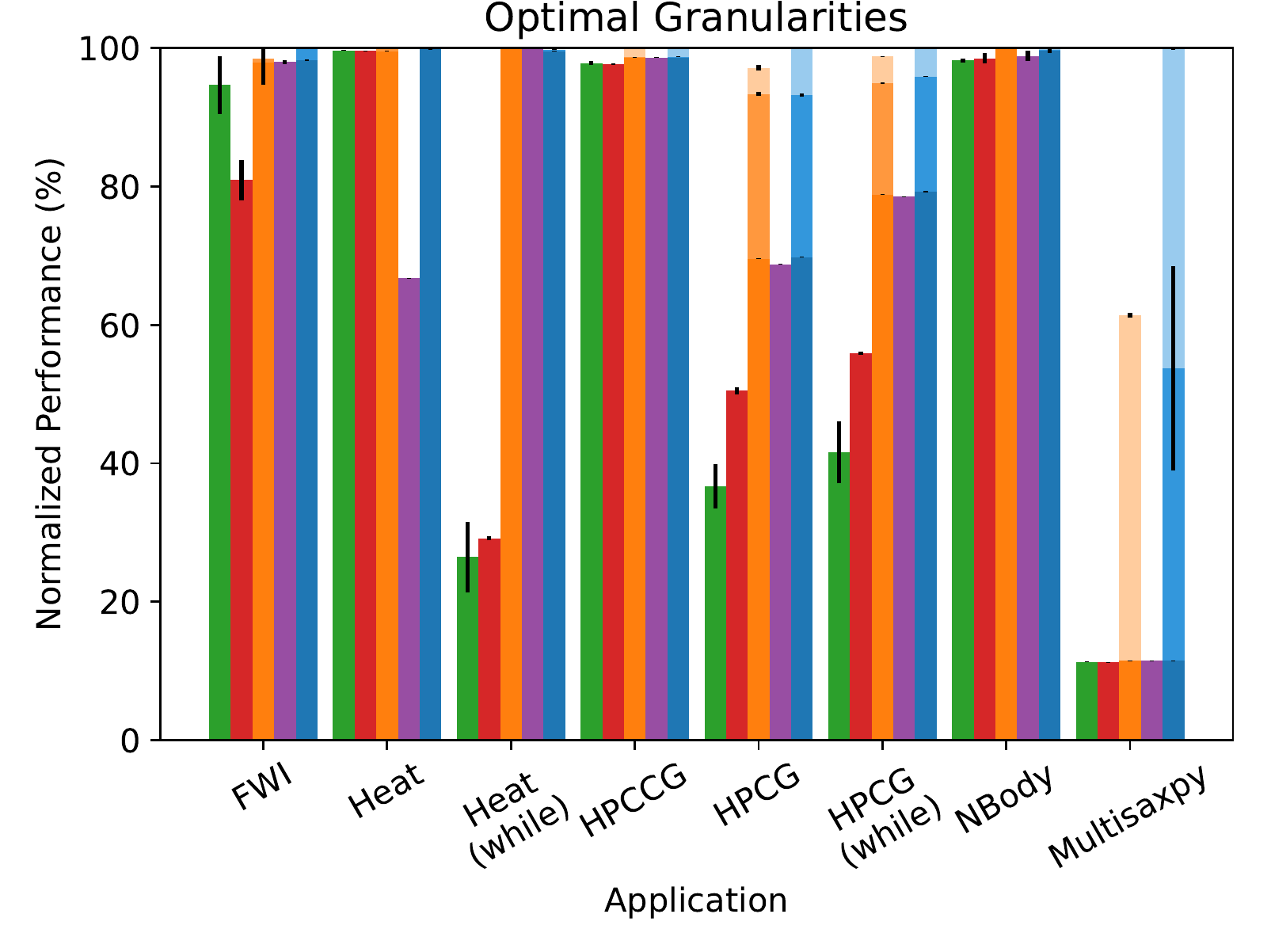}
		\caption{Optimal granularities}
		\label{fig:perf-peak}
	\end{subfigure}
	\begin{subfigure}[b]{.48\textwidth}
		\centering
		\includegraphics[width=\columnwidth]{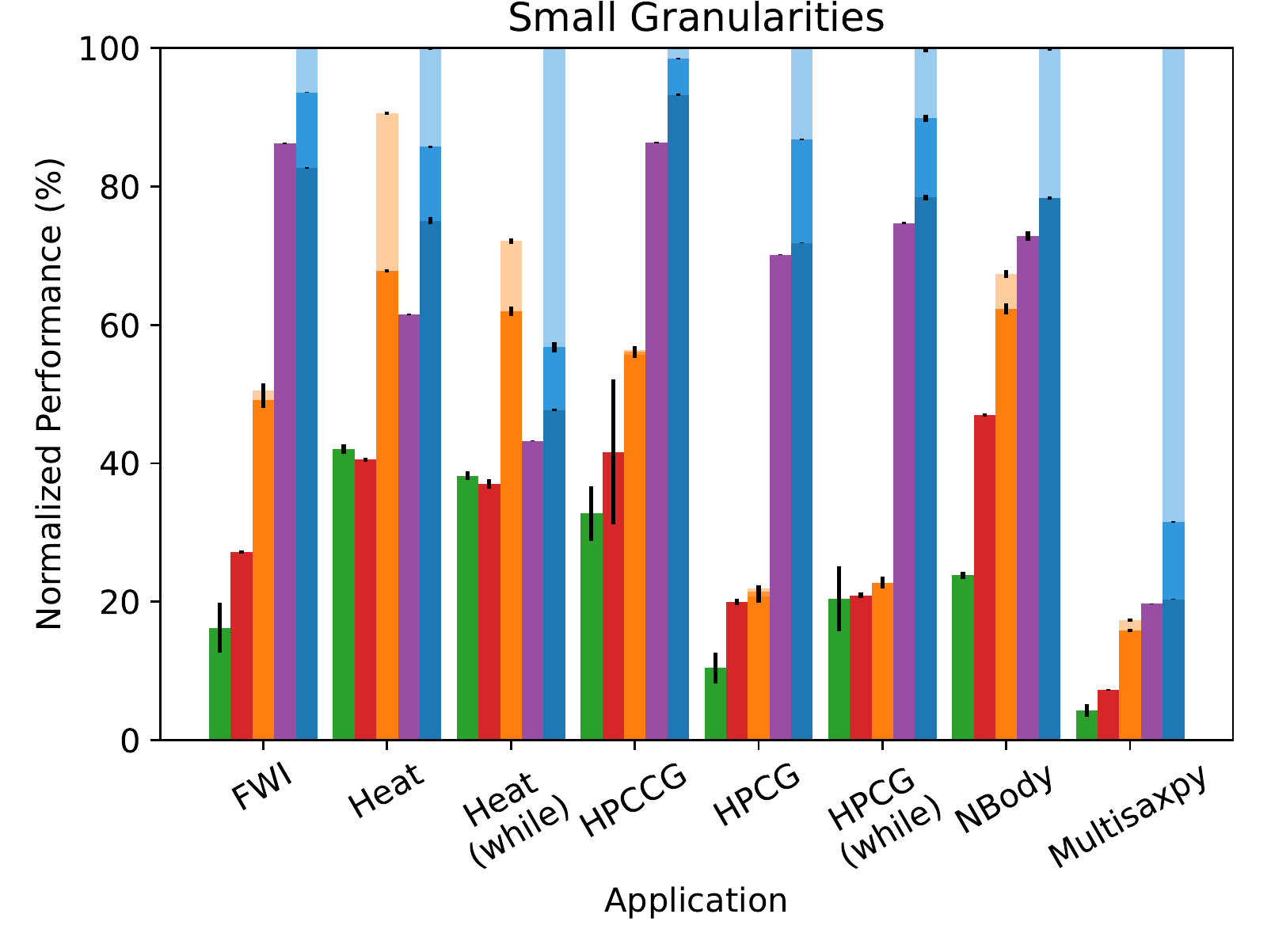}
		\caption{Small granularities}
		\label{fig:perf-small}
	\end{subfigure}
	\caption{Performance comparison of different variants when executed with optimal and small granularities}
\end{figure*}

\begin{figure*}[ht!]
	\centering
	\begin{subfigure}[b]{.01\textwidth}
		\centering
		\hspace{1em}
	\end{subfigure}
	\hfill
	\begin{subfigure}[b]{.9\textwidth}
		\centering
		\includegraphics[width=\textwidth]{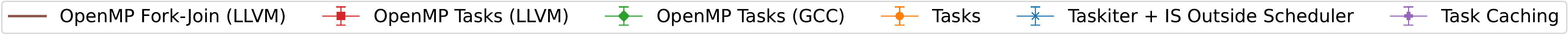}
	\end{subfigure}
	\hfill
	\begin{subfigure}[b]{.01\textwidth}
		\centering
		\hspace{1em}
	\end{subfigure}
	\begin{subfigure}[b]{0.24\textwidth}
		\centering
		\includegraphics[width=\textwidth]{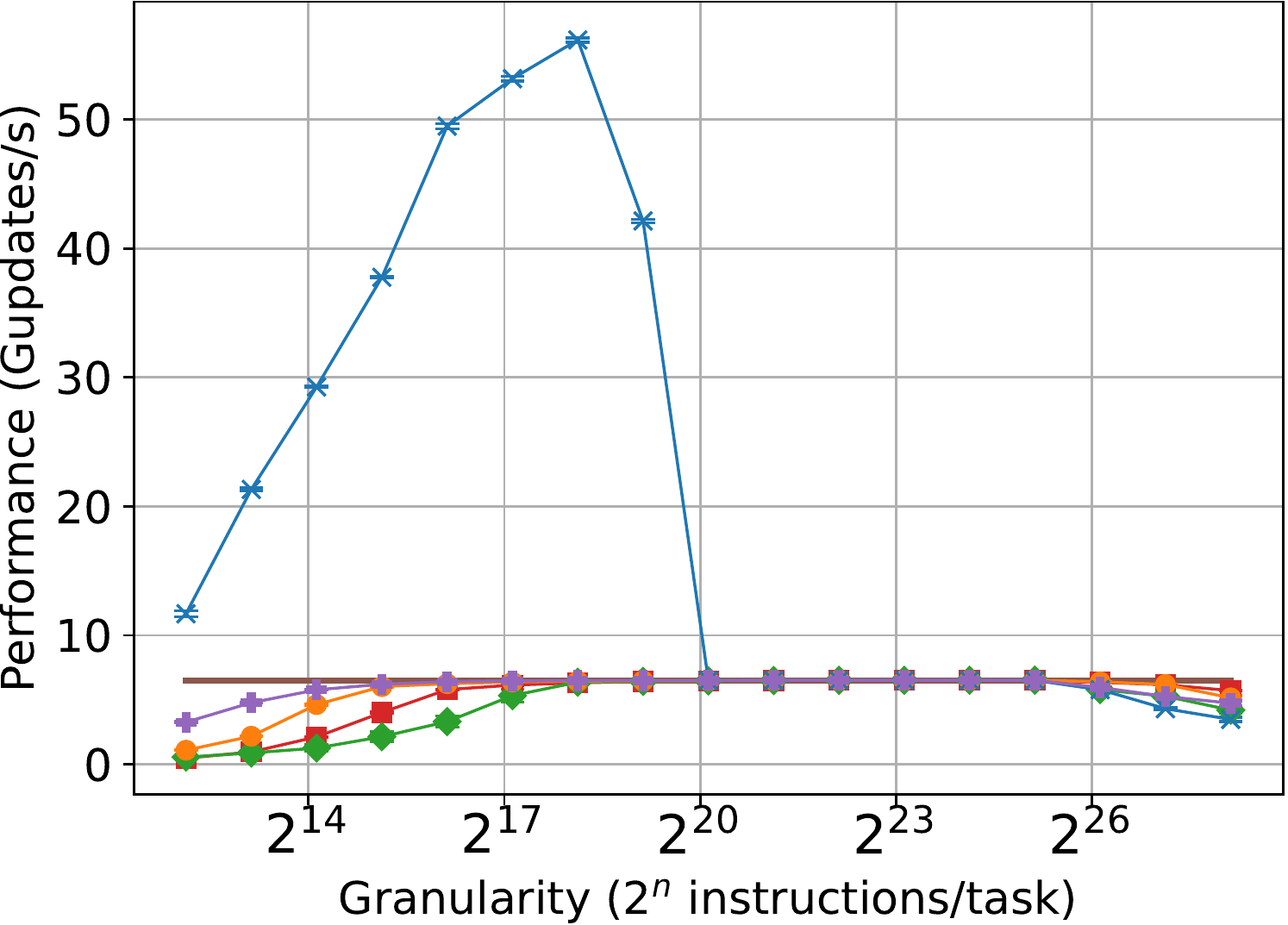}
		\caption{Multisaxpy}
		\label{fig:exp:saxpy}
	\end{subfigure}
	\hfill
	\begin{subfigure}[b]{0.24\textwidth}
		\centering
		\includegraphics[width=\textwidth]{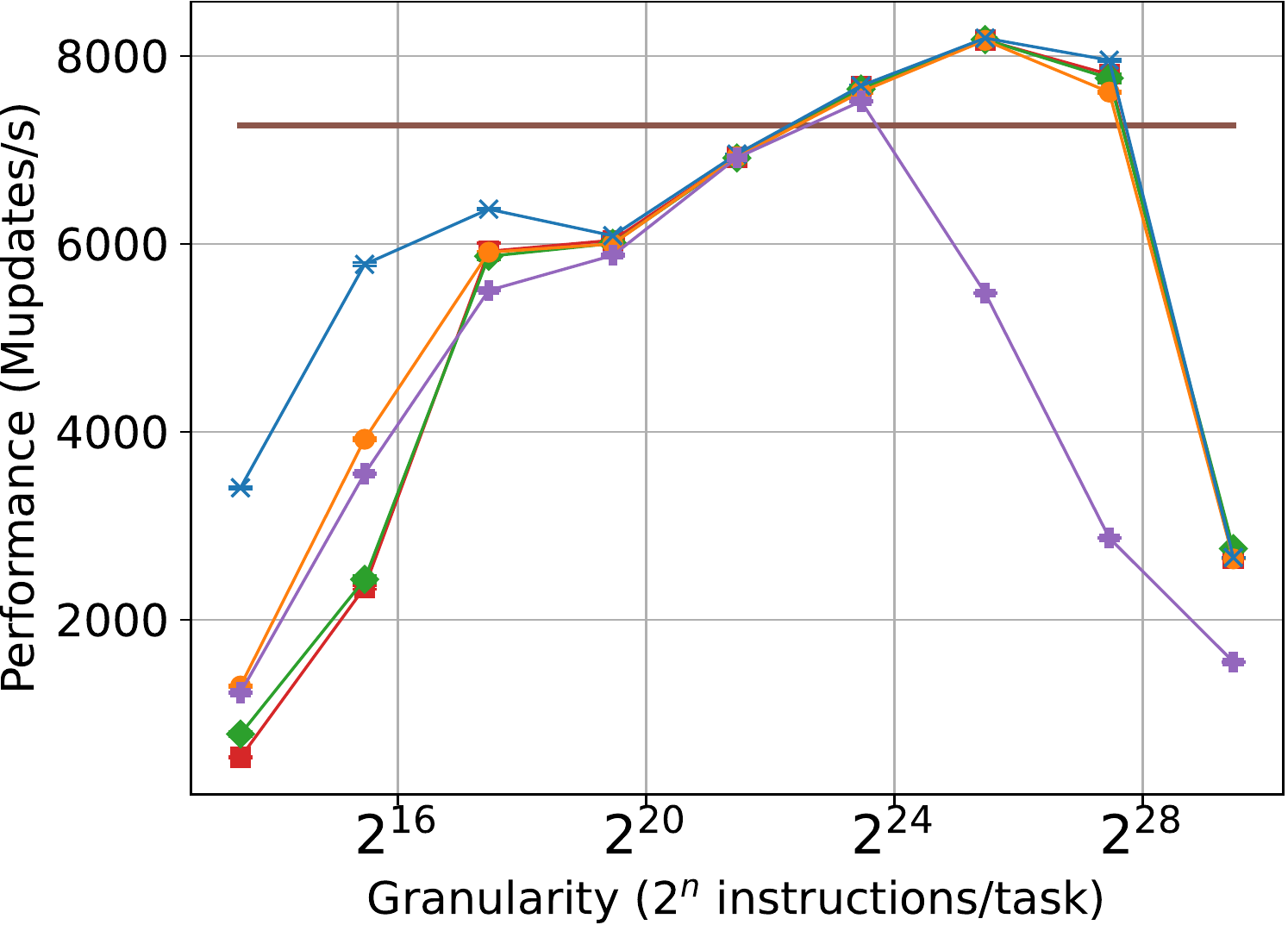}
		\caption{Gauss-Seidel}
		\label{fig:exp:heat}
	\end{subfigure}
	\hfill
	\begin{subfigure}[b]{0.24\textwidth}
		\centering
		\includegraphics[width=\textwidth]{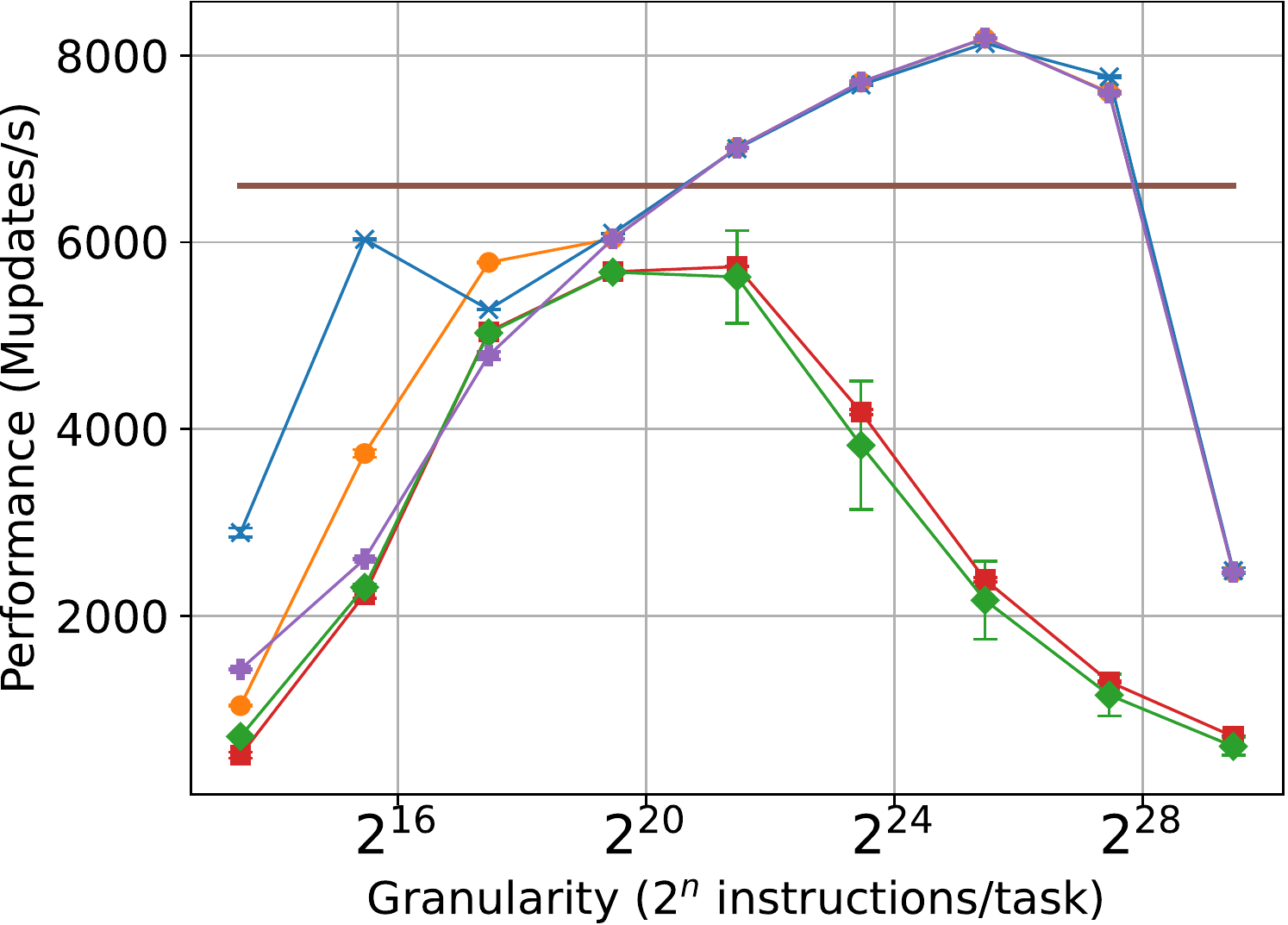}
		\caption{Gauss-Seidel (while)}
		\label{fig:exp:heat-while}
	\end{subfigure}
	\hfill
	\begin{subfigure}[b]{0.24\textwidth}
		\centering
		\includegraphics[width=\textwidth]{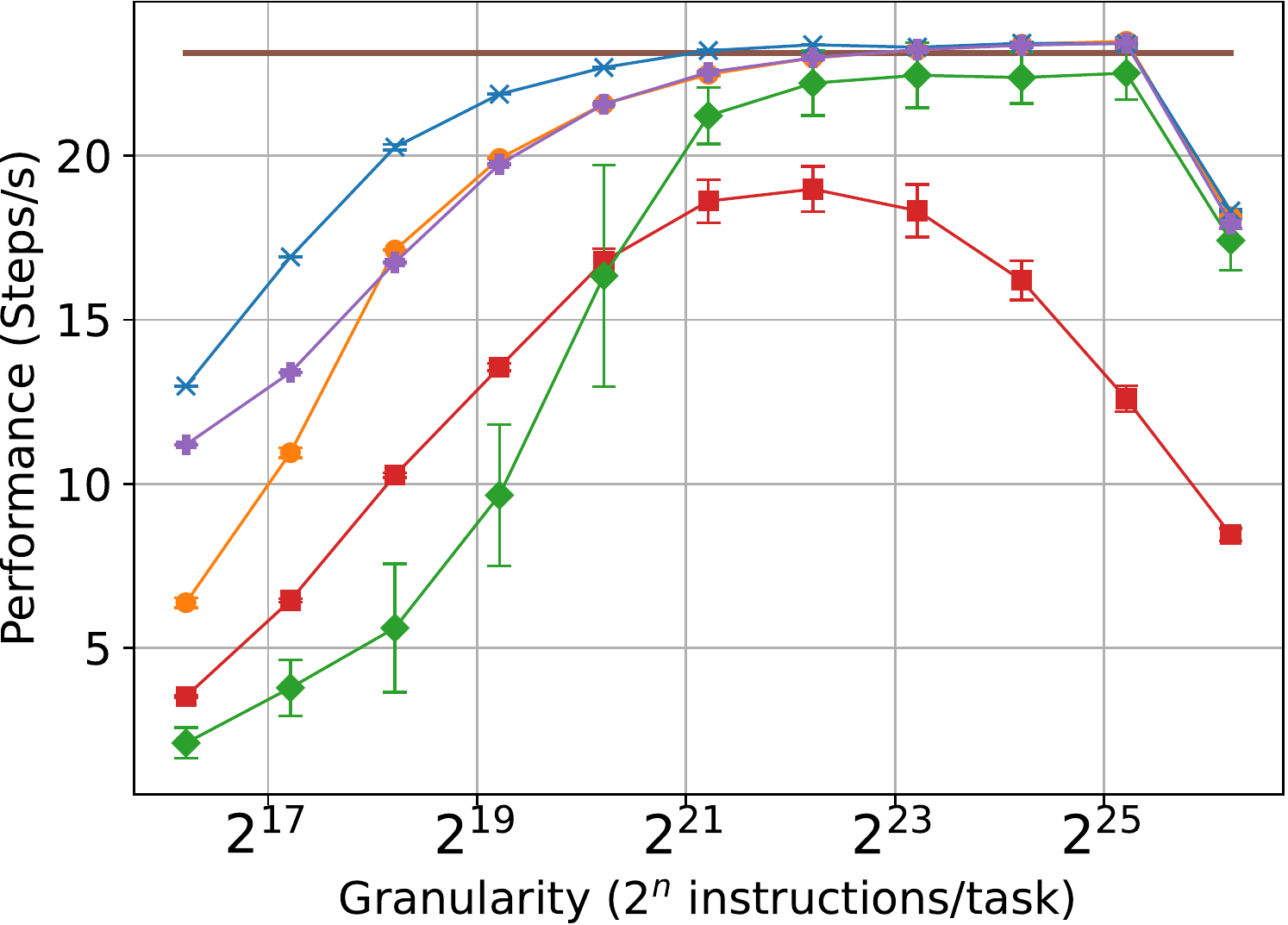}
		\caption{Full-Waveform Inversion}
		\label{fig:exp:fwi}
	\end{subfigure}
	\hfill
	\begin{subfigure}[b]{0.24\textwidth}
		\vspace{1em}
		\centering
		\includegraphics[width=\textwidth]{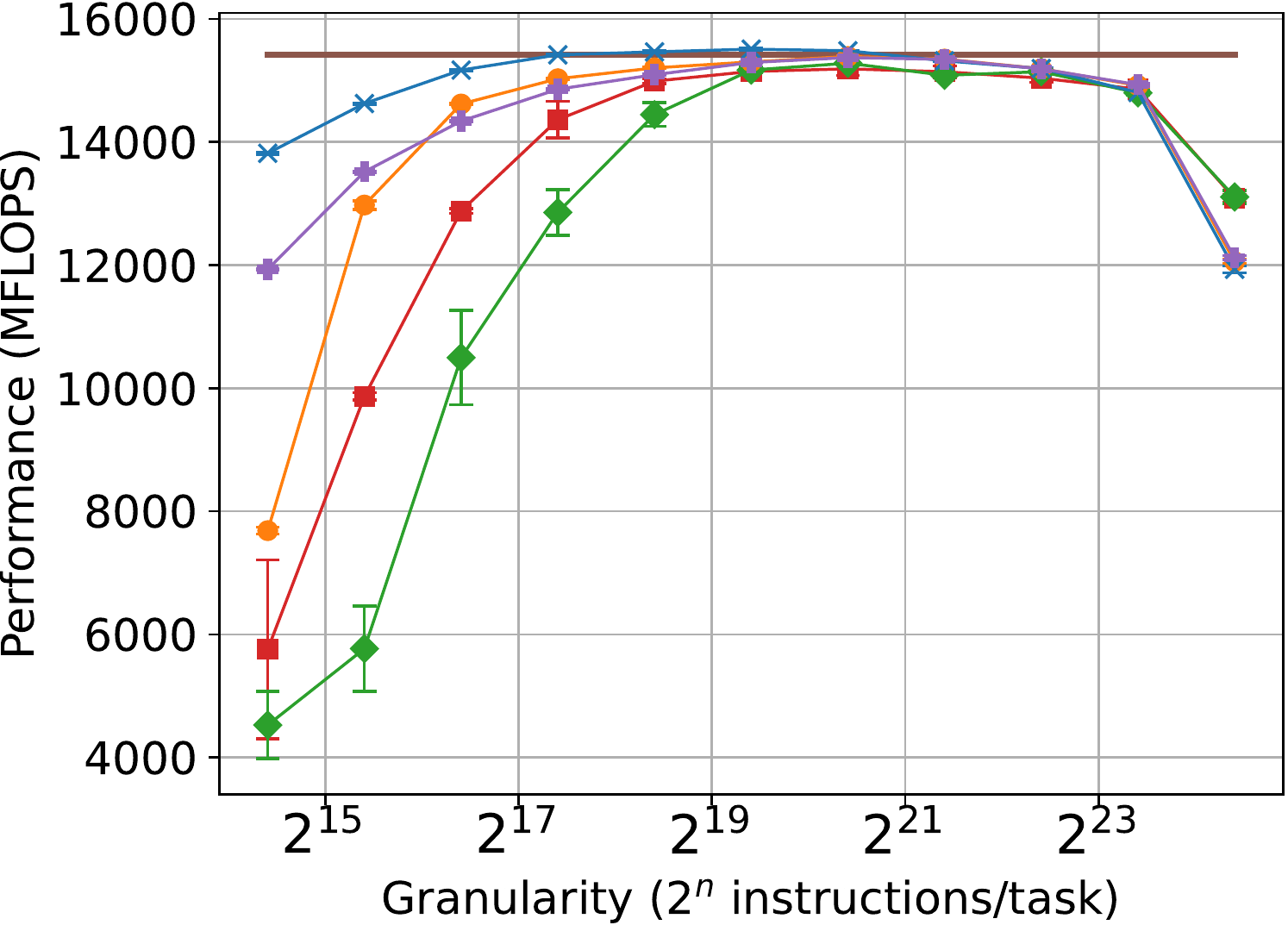}
		\caption{HPCCG}
		\label{fig:exp:hpccg}
	\end{subfigure}
	\hfill
	\begin{subfigure}[b]{0.24\textwidth}
		\centering
		\includegraphics[width=\textwidth]{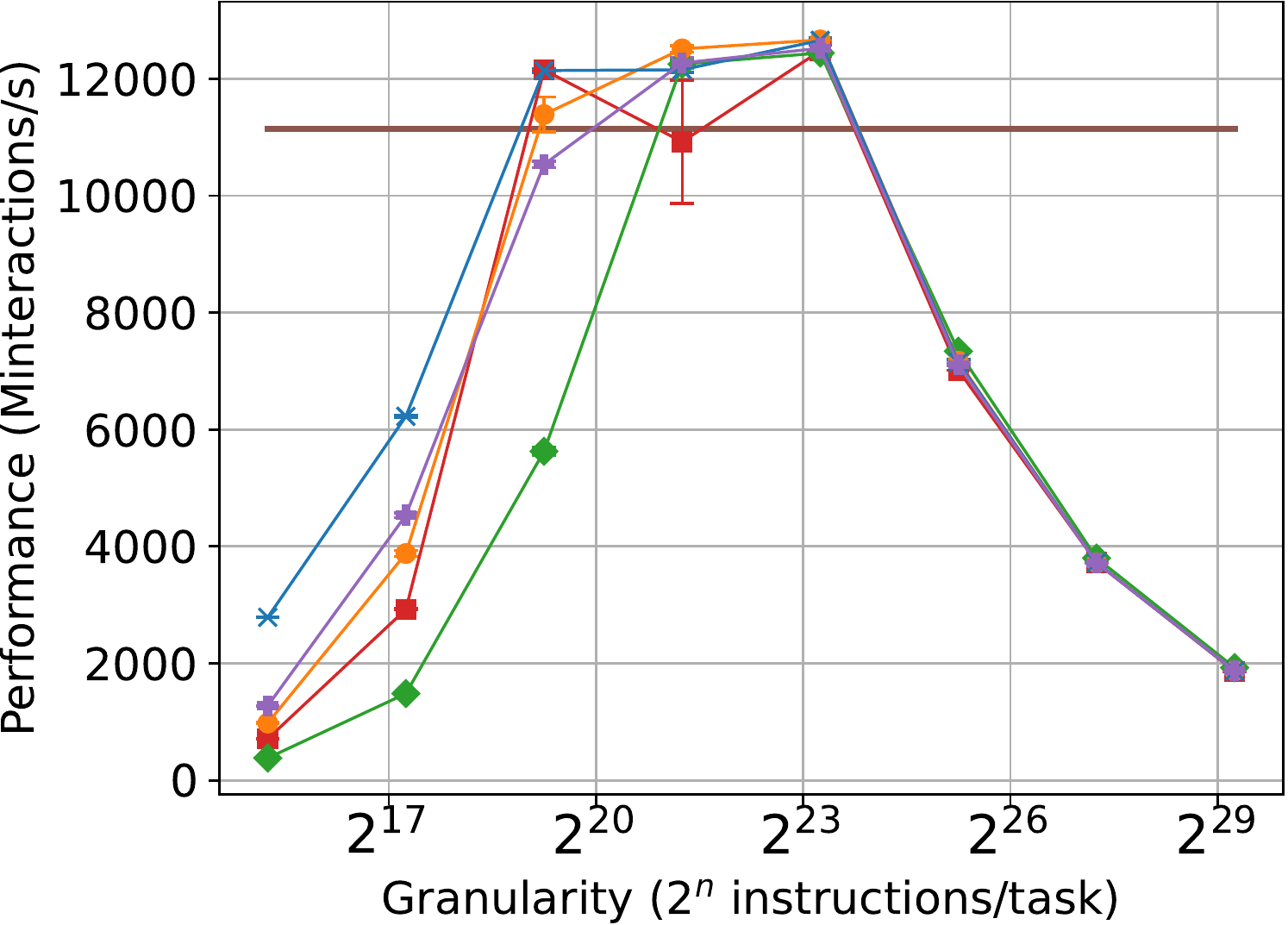}
		\caption{N-Body}
		\label{fig:exp:nbody}
	\end{subfigure}
	\hfill
	\begin{subfigure}[b]{0.24\textwidth}
		\centering
		\includegraphics[width=\textwidth]{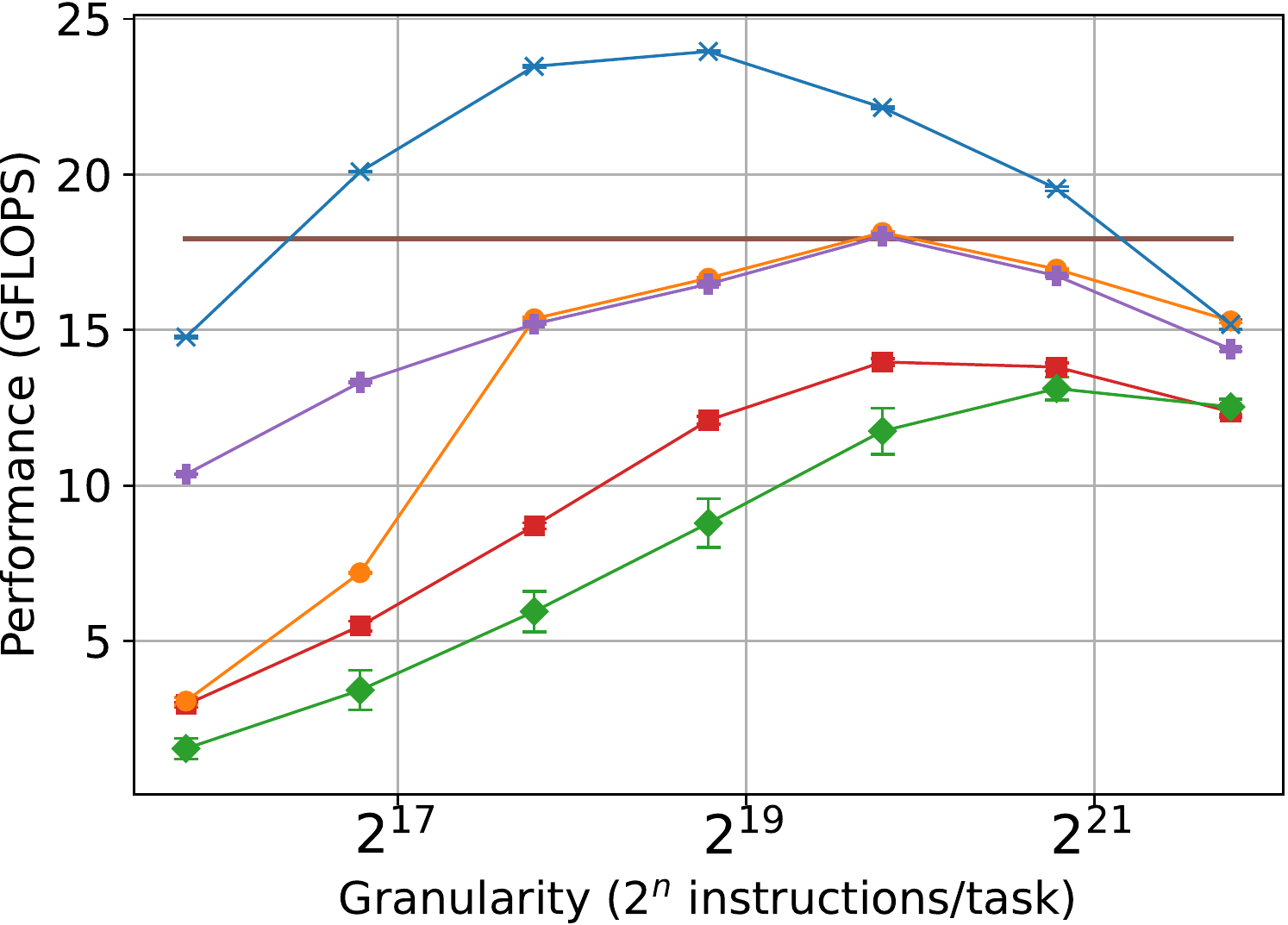}
		\caption{HPCG}
		\label{fig:exp:hpcg}
	\end{subfigure}
	\hfill
	\begin{subfigure}[b]{0.24\textwidth}
		\vspace{1em}
		\centering
		\includegraphics[width=\textwidth]{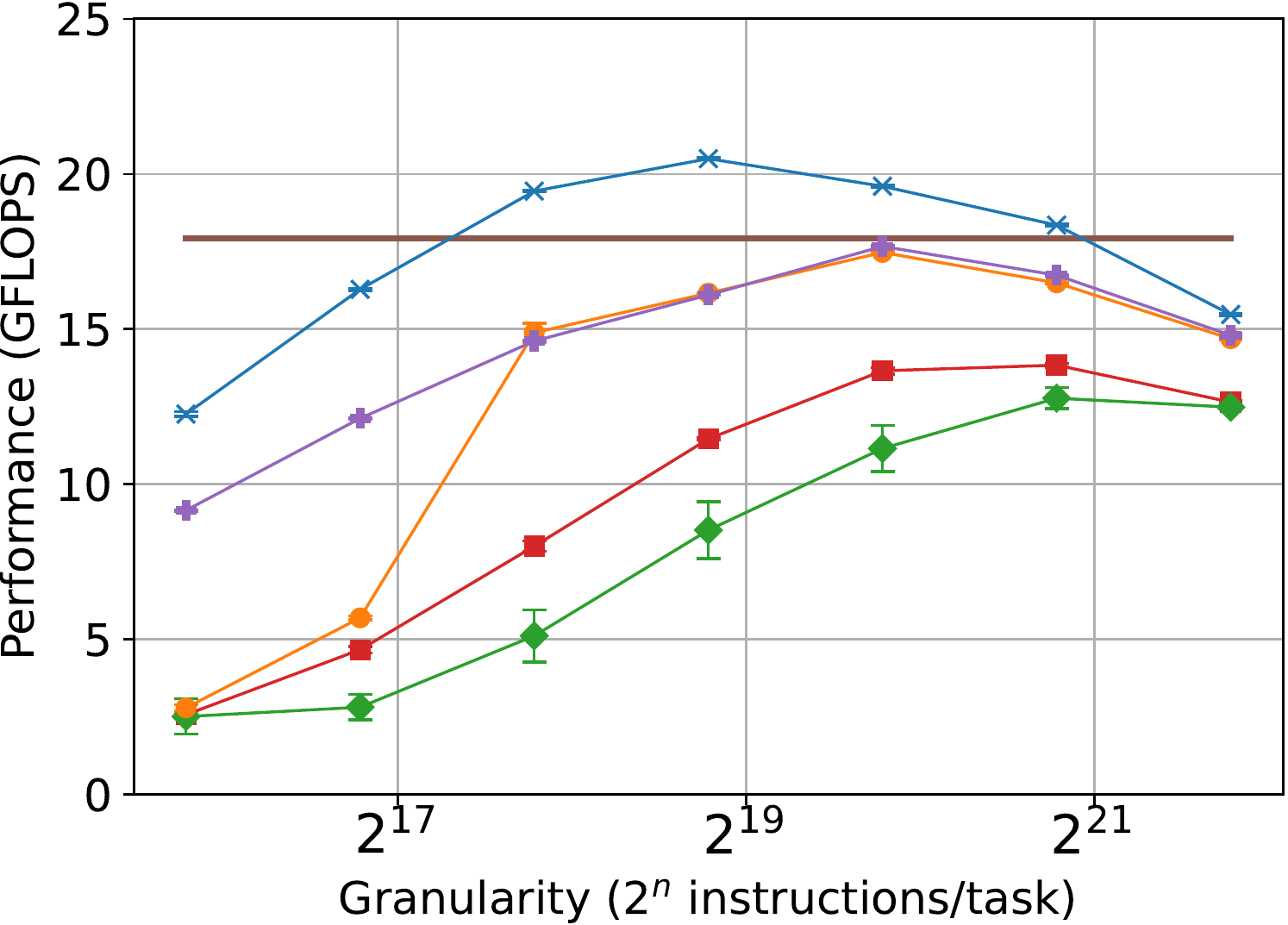}
		\caption{HPCG (while)}
		\label{fig:exp:hpcg-normr}
	\end{subfigure}
	\caption{Extended experimental evaluation for all granularities on the Taskiter with Immediate Successor against OpenMP}
	\label{fig:exp}
\end{figure*}

In the first experiment, we measure the \textit{normalized performance}, which is the performance of a specific execution relative to the maximum performance of all executions.
This normalized performance is obtained based on the figure of merit provided by each application,
and absolute performance figures for all granularities are presented later in Section \ref{sec:second-exp}.
We run the experiment on two different configurations to observe the most relevant task granularities:
First, at the optimal granularity, where performance is in the optimal region.
Second, at a point where tasks are too small but still more than $50\%$ peak performance is achieved.
This second scenario simulates a strong scaling situation, thus evaluating the scalability of each solution.

In this experiment, we test seven variants of the OmpSs-2 programming model to verify the effect of both the taskiter and the immediate successor policy:
\begin{enumerate}
	\item \textit{Tasks} is the base OmpSs-2 version of the application, using tasks with dependencies, and no immediate successor.
	\item \textit{Tasks + Immediate Successor (IS)} is the same as the \textit{Tasks} version but applies the immediate successor policy.
	However, this immediate successor is only applied inside the synchronization mechanism for the task scheduler.
	\item \textit{Tasks + IS Outside Scheduler} is the \textit{Tasks} version using the immediate successor policy and bypassing the scheduler when possible.
	\item \textit{Task Caching} is the application adapted to simulate a task caching approach. We implemented the semantics of the \texttt{taskgraph} construct~\cite{taskgraph},
	where all data structures are cached between iterations, but without transformation or matching
	of dependencies between one iteration and the next.
	\item \textit{Taskiter} is the application adapted to use a taskiter to transform the main loop into a cyclic graph.
	\item \textit{Taskiter + Immediate Successor (IS)} is the same as the \textit{Taskiter} version but applies the immediate successor policy inside the
	scheduler's synchronization mechanism.
	\item \textit{Taskiter + IS Outside Scheduler} is the \textit{Taskiter} version using the immediate successor policy and bypassing the scheduler when possible.
\end{enumerate}

Note that we split the evaluation for the IS policy into two parts.
First, we apply the logic behind the immediate successor policy, but every task still has to go through the existing scheduler queues (the \textit{Immediate Successor} version).
This way, we can measure when performance increases thanks to better data locality instead of just the reduction of scheduling overhead.
In the second part, we also use the immediate successor to bypass the scheduler altogether when an appropriate candidate is found, reducing the contention
in the scheduler (the \textit{IS Outside Scheduler} version).

Additionally, we compare every result with an equivalent OpenMP tasks version of the application. We used two different runtimes for the comparison:
the GOMP runtime provided by GCC 10.2.0 and the LLVM OpenMP Runtime on its 13.0.0-rc1 version.
We chose to compare against the GCC runtime as a reference implementation for OpenMP, and against the LLVM runtime because it is based on the Intel OpenMP runtime,
which is known to have very competitive performance.

Figure \ref{fig:perf-peak} shows the performance of each evaluated version versus the maximum figure of merit for each benchmark.
Note that we stack the improvements of the immediate successor policies. For instance, the solid orange color bar refers to the
\textit{Tasks} version, while lighter orange bars show how the \textit{Tasks} version performs when combined with the two immediate
successor policies.

Generally, reducing overheads should have little effect on optimal granularities, but any locality improvements may be noticeable.
There are several key insights we can extract from these results.
First, if we focus exclusively on the \textit{Tasks} versus the \textit{Taskiter} version, we observe
that performance remains very similar. This is expected as these runs happen with optimal granularities,
and task creation overhead does not limit performance.
However, when we introduce the immediate successor policy, we can achieve higher peak performance on the HPCCG,
HPCG and multisaxpy benchmarks, thanks to increased data locality.
Moreover, there is a significant difference between placing the IS policy inside and outside the scheduler locking system,
suggesting that scheduling overhead plays a prominent role even in optimal granularities.

For all benchmarks, the performance of the studied versions is either competitive or superior to OpenMP.

Figure \ref{fig:perf-small} shows the same results for smaller granularities, where we measure the scalability of each version.
Again, we normalized the performance to that of the best-performing version.
In this case, the \textit{Tasks} version always performs better than its OpenMP equivalents, which confirms that our baseline is already
a very scalable runtime.
In turn, the \textit{Taskiter} version shows a better performance for small granularities than \textit{Tasks}, thanks to its
reduced task creation and dependency management overheads.
We find that the most scalable and best-performing version is the \textit{Taskiter + IS Outside Scheduler}.

We can also observe the synergistic effects of both contributions for small granularities.
For example, in the HPCG benchmark, the IS policy produces no performance improvement for the \textit{Tasks} version
but strongly affects the \textit{Taskiter} version's performance. This is observed in many benchmarks,
where the \textit{Tasks + IS Outside Scheduler} has a much smaller speedup than the \textit{Taskiter + IS Outside Scheduler} version due
to the synergy between both contributions.
We analyze further the HPCG's case in Section \ref{sec:traces}.

In optimal granularities, the task caching approach either works similarly to the base \textit{Tasks} and \textit{Taskiter} version or causes a slowdown
in cases like the Heat equation, where adding a barrier between iterations decreases the available parallelism due to its wave-front pattern.
In small granularities, the performance of task caching generally sits between the \textit{Tasks} and \textit{Taskiter} versions as a middle-ground.
However, it is outperformed by the construct proposed in this paper in all experiments.

\begin{figure*}[ht!]
	\centering
	\begin{subfigure}[b]{1\textwidth}
		\centering
		\includegraphics[width=.65\textwidth]{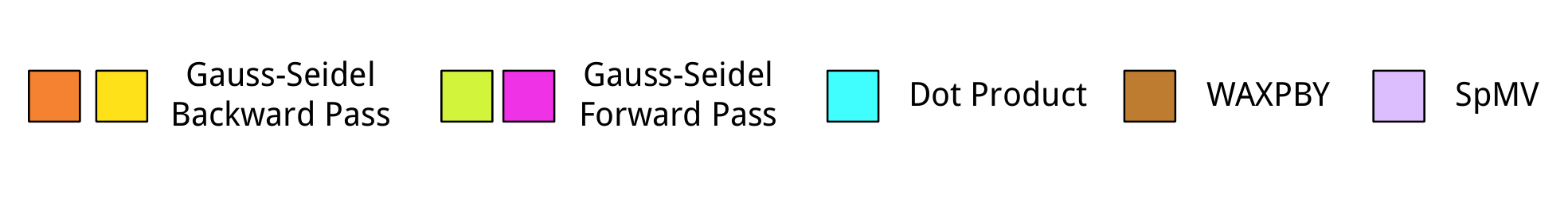}
	\end{subfigure}
	\begin{subfigure}[b]{1\textwidth}
		\centering
		\includegraphics[width=\textwidth]{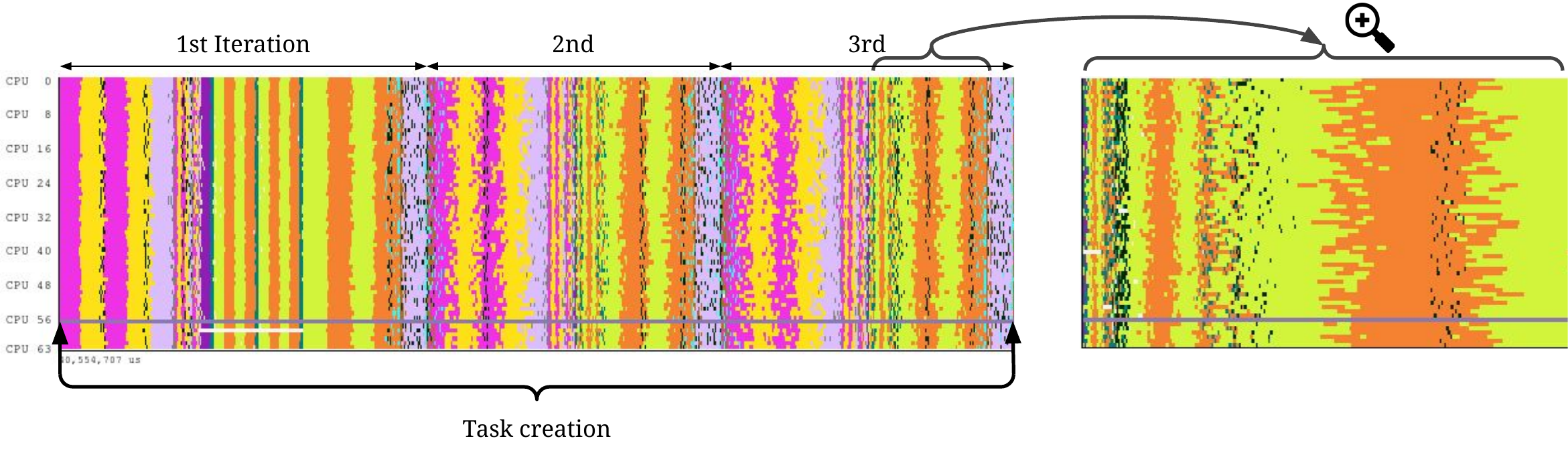}
		\caption{Version parallelized using OmpSs-2 tasks}
		\label{fig:traces:std}
	\end{subfigure}
	\begin{subfigure}[b]{1\textwidth}
		\centering
		\includegraphics[width=\textwidth]{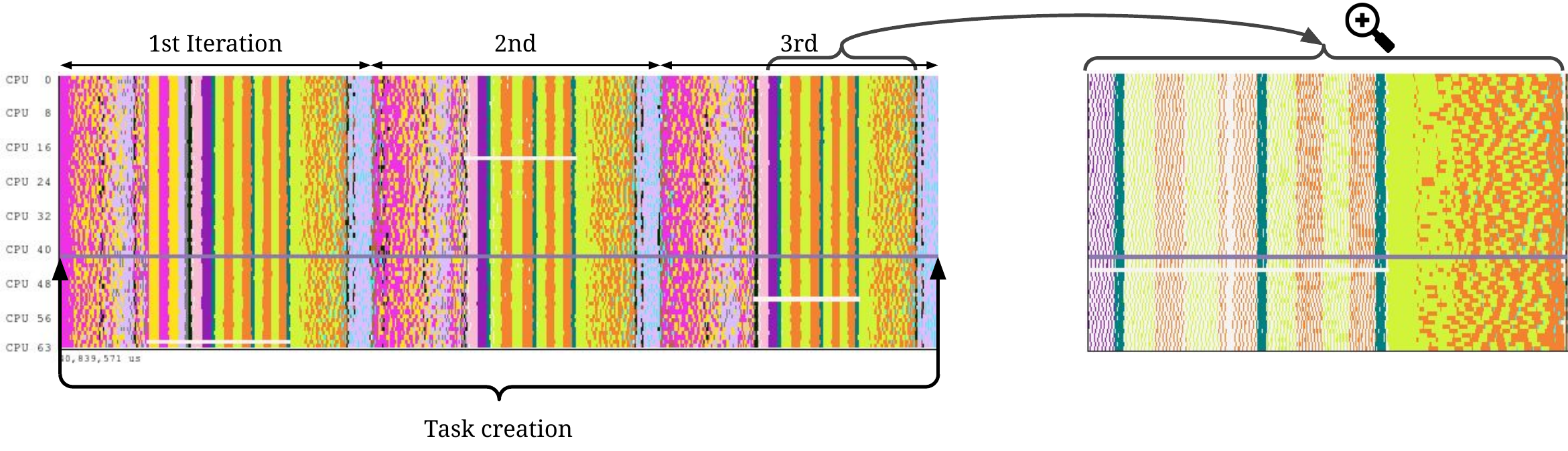}
		\caption{Version parallelized using OmpSs-2 tasks and the immediate successor policy}
		\label{fig:traces:std-imm}
	\end{subfigure}
	\begin{subfigure}[b]{1\textwidth}
		\centering
		\includegraphics[width=\textwidth]{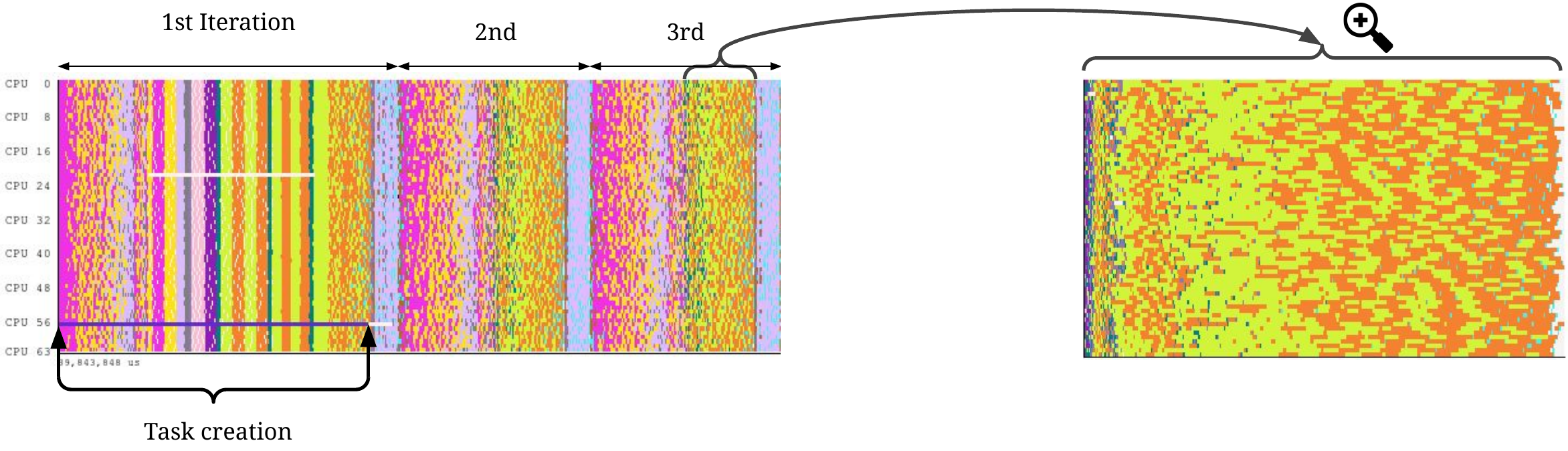}
		\caption{Version parallelized using the taskiter construct and the immediate successor policy}
		\label{fig:traces:tg-imm}
	\end{subfigure}
	\caption{Execution traces for the first three iterations of the HPCG benchmark using 64 cores of the AMD machine. The time scale of each trace is the same, a shorter trace implies less elapsed time. }
	\label{fig:traces}
\end{figure*}

\subsection{Experiment 2: Comparison against the baseline}
\label{sec:second-exp}

The second experiment evaluates performance on a more extensive range of
task granularities for the \textit{Taskiter + IS Outside Scheduler} version against both baseline OpenMP task implementations
and a fork-join OpenMP implementation.
The fork-join implementation was done using \texttt{omp for} constructs on the parallelizable
parts of each benchmark, without major code changes except for the Gauss-Seidel,
where the code was adapted to iterate on the matrix's elements diagonally so the loop could be parallelized.

With this comparison, we want to show how the proposed improvements affect the scalability of task-based applications.
Moreover, we will show how the changes can make task-based programs compete and outperform worksharing.

Overall, results from Figure \ref{fig:exp} show that both the taskiter and the immediate successor policy improve the performance
of task-based iterative applications. We measure task granularity on \textit{instructions} per task, as it is a metric that does not vary between
different versions and directly correlates with time. Note that fork-join versions do not have task granularity, and hence are represented as straight lines.

One result that stands out is Figure~\ref{fig:exp:saxpy}, where there is a very notable performance difference
for specific granularities in the Multisaxpy benchmark.
In this benchmark, starting at a granularity of $2^{19}$, the working set of a task fits into its L3 cache slice.
Therefore, if another task using the same data is immediately scheduled into the same core, all of its working set is hot in cache,
which is precisely what the immediate successor policy does.
Moreover, this effect synergizes with the taskiter, as tasks are created once and scheduled multiple times in the same core.
On the other hand, the OpenMP runtimes do not have this locality policy implemented and schedule other tasks instead, and the
worksharing version has to finish one iteration before the next one starts.
We achieve an 8.75x speedup in the optimal granularity compared to the OpenMP baseline.

Another relevant insight is the Gauss-Seidel heat equation in Figures~\ref{fig:exp:heat} and~\ref{fig:exp:heat-while}.
In the first variant, where the number of iterations is fixed, the peak performance obtained is the same for all runtimes,
and the difference is seen in the smaller granularities only.
However, when we check the residual at each iteration, the performance of OpenMP drops due to the barriers introduced by task reductions.
In OmpSs-2, reductions do not imply barriers, and the implementation of the taskiter allows to overlap execution of tasks belonging to different iterations, maintaining the available parallelism.
This is also the reason that the taskiter is the only version able to outperform fork-join parallelism in Figure~\ref{fig:exp:heat-while}.

Moreover, our proposals are also able to outperform fork-join parallelism and OpenMP tasks in the HPCG benchmark shown in Figures~\ref{fig:exp:hpcg} and~\ref{fig:exp:hpcg-normr}.

\subsection{HPCG Execution Traces}
\label{sec:traces}

So far, we have seen the performance improvements that both the taskiter and the immediate successor policy can deliver.
However, we can also leverage the instrumentation included in OmpSs-2 to obtain execution traces and study exactly
how our contributions affect each application.
We chose to study the HPCG benchmark, which is affected by both contributions and showcases its synergistic effects.
We obtained execution traces of the application for the granularity highlighted in Figure \ref{fig:exp:hpcg},
and we show these traces in Figure \ref{fig:traces}.
In all the traces, each row represents one of the 64 cores of the machine. The $x$-axis represents time, and
each color is a different \textit{task type}, which we use to identify different phases of the application.
The time scale of each trace is the same, but only three iterations are shown. For each trace, we provide a not to scale zoomed section of a small subset of the execution.

The first trace displayed in Figure \ref{fig:traces:std} shows the baseline OmpSs-2 tasks version of the HPCG benchmark.
Colors denote tasks from different application phases, revealing an iterative pattern.
Arrows below the trace highlight the task creator core. This thread executes the \textit{main} task, which creates all other tasks
to be executed by the rest of the cores.

When we introduce the immediate successor policy, as seen in Figure \ref{fig:traces:std-imm}, task creation remains the same,
but task scheduling changes. In contrast to the well-defined phases on the previous trace, tasks are instead executed in a different order in some instances (see the zoomed-in section).
This happens because each orange task depends on a yellow task, and the immediate successor policy decides to schedule one after the other. Note that sections that display this pattern are \textit{shorter} than in the previous trace because
better data locality causes tasks to execute faster, as part of the working set is hot in the cache.
Moreover, as shown in the zoomed-in section, this produces an unexpected side-effect. As tasks execute faster,
the task creator cannot keep up and fails to create tasks fast enough to feed all the cores, producing a starvation scenario.

The taskiter solves this starvation problem in Figure \ref{fig:traces:tg-imm}.
In this case, the task creation is done only during the first iteration, and then a DCTG is constructed, and there is no need to create tasks again.
The first iteration is as slow as the other cases, but the following iterations are much shorter because task creation does not limit
performance. Moreover, as all tasks are created, we can apply the immediate successor policy more effectively, maximizing locality and exploiting
the memory hierarchy better.

\section{Conclusions}
\label{sec:conclusions}

In this work, we have presented a new directive for OmpSs-2 and OpenMP, the \textit{taskiter}.
We have shown how it fits naturally into iterative HPC applications and delivers significant performance
gains thanks to reducing task creation and dependency management overheads.
Moreover, we have combined it with a scalable and straightforward immediate successor heuristic that preserves data locality
while reducing scheduling overheads.

Our evaluation shows that applying both techniques to task-based iterative applications delivers significant
scalability improvements and speedups, achieving an average speedup of 3.7x for small granularities compared to the reference OmpSs-2 implementation
and a 5x and 7.46x speedup over the LLVM and GCC OpenMP runtimes, respectively.
Moreover, the resulting task-based applications using the right granularity can compete or outperform worksharing versions of all benchmarks.

In future work, we plan to extend the \textit{taskiter} construct to support device tasks in heterogeneous applications to expand its applicability further.

\bibliographystyle{ACM-Reference-Format}
\bibliography{paper}


\begin{thebibliography}{27}


\ifx \showCODEN    \undefined \def \showCODEN     #1{\unskip}     \fi
\ifx \showDOI      \undefined \def \showDOI       #1{#1}\fi
\ifx \showISBNx    \undefined \def \showISBNx     #1{\unskip}     \fi
\ifx \showISBNxiii \undefined \def \showISBNxiii  #1{\unskip}     \fi
\ifx \showISSN     \undefined \def \showISSN      #1{\unskip}     \fi
\ifx \showLCCN     \undefined \def \showLCCN      #1{\unskip}     \fi
\ifx \shownote     \undefined \def \shownote      #1{#1}          \fi
\ifx \showarticletitle \undefined \def \showarticletitle #1{#1}   \fi
\ifx \showURL      \undefined \def \showURL       {\relax}        \fi
\providecommand\bibfield[2]{#2}
\providecommand\bibinfo[2]{#2}
\providecommand\natexlab[1]{#1}
\providecommand\showeprint[2][]{arXiv:#2}

\bibitem[\protect\citeauthoryear{Akhmetova, Kestor, Gioiosa, Markidis, and
  Laure}{Akhmetova et~al\mbox{.}}{2015}]%
        {GranularityAnalysis2015}
\bibfield{author}{\bibinfo{person}{Dana Akhmetova}, \bibinfo{person}{Gokcen
  Kestor}, \bibinfo{person}{Roberto Gioiosa}, \bibinfo{person}{Stefano
  Markidis}, {and} \bibinfo{person}{Erwin Laure}.}
  \bibinfo{year}{2015}\natexlab{}.
\newblock \showarticletitle{On the Application Task Granularity and the
  Interplay with the Scheduling Overhead in Many-Core Shared Memory Systems}.
  In \bibinfo{booktitle}{\emph{2015 IEEE International Conference on Cluster
  Computing}}. \bibinfo{pages}{428--437}.
\newblock
\urldef\tempurl%
\url{https://doi.org/10.1109/CLUSTER.2015.65}
\showDOI{\tempurl}


\bibitem[\protect\citeauthoryear{\'{A}lvarez, Sala, Maro\~{n}as, Roca, and
  Beltran}{\'{A}lvarez et~al\mbox{.}}{2021}]%
        {advancedsynchronizationtechniques}
\bibfield{author}{\bibinfo{person}{David \'{A}lvarez}, \bibinfo{person}{Kevin
  Sala}, \bibinfo{person}{Marcos Maro\~{n}as}, \bibinfo{person}{Aleix Roca},
  {and} \bibinfo{person}{Vicen\c{c} Beltran}.} \bibinfo{year}{2021}\natexlab{}.
\newblock \showarticletitle{Advanced Synchronization Techniques for Task-Based
  Runtime Systems}. In \bibinfo{booktitle}{\emph{Proceedings of the 26th ACM
  SIGPLAN Symposium on Principles and Practice of Parallel Programming}}
  (Virtual Event, Republic of Korea) \emph{(\bibinfo{series}{PPoPP '21})}.
  \bibinfo{publisher}{Association for Computing Machinery},
  \bibinfo{address}{New York, NY, USA}, \bibinfo{pages}{334–347}.
\newblock
\showISBNx{9781450382946}
\urldef\tempurl%
\url{https://doi.org/10.1145/3437801.3441601}
\showDOI{\tempurl}


\bibitem[\protect\citeauthoryear{Balart, Duran, Gon, Martorell, Ayguadé, and
  Labarta}{Balart et~al\mbox{.}}{2004}]%
        {mercurium2}
\bibfield{author}{\bibinfo{person}{J Balart}, \bibinfo{person}{A Duran},
  \bibinfo{person}{Mg Gon}, \bibinfo{person}{X Martorell}, \bibinfo{person}{E
  Ayguadé}, {and} \bibinfo{person}{Jesús Labarta}.}
  \bibinfo{year}{2004}\natexlab{}.
\newblock \showarticletitle{Nanos mercurium: A research compiler for OpenMP}.
\newblock \bibinfo{journal}{\emph{Proceedings of the European Workshop on
  OpenMP}}  \bibinfo{volume}{8} (\bibinfo{date}{01} \bibinfo{year}{2004}).
\newblock


\bibitem[\protect\citeauthoryear{Berger, McKinley, Blumofe, and Wilson}{Berger
  et~al\mbox{.}}{2000}]%
        {hoard}
\bibfield{author}{\bibinfo{person}{Emery~D. Berger},
  \bibinfo{person}{Kathryn~S. McKinley}, \bibinfo{person}{Robert~D. Blumofe},
  {and} \bibinfo{person}{Paul~R. Wilson}.} \bibinfo{year}{2000}\natexlab{}.
\newblock \showarticletitle{Hoard: A Scalable Memory Allocator for
  Multithreaded Applications}. In \bibinfo{booktitle}{\emph{Proceedings of the
  Ninth International Conference on Architectural Support for Programming
  Languages and Operating Systems}} (Cambridge, Massachusetts, USA)
  \emph{(\bibinfo{series}{ASPLOS IX})}. \bibinfo{publisher}{Association for
  Computing Machinery}, \bibinfo{address}{New York, NY, USA},
  \bibinfo{pages}{117–128}.
\newblock
\showISBNx{1581133170}
\urldef\tempurl%
\url{https://doi.org/10.1145/378993.379232}
\showDOI{\tempurl}


\bibitem[\protect\citeauthoryear{BSC}{BSC}{2020}]%
        {bsc2020ompss2}
\bibfield{author}{\bibinfo{person}{BSC}.} \bibinfo{year}{2020}\natexlab{}.
\newblock \bibinfo{title}{{OmpSs-2} Specification}.
\newblock
\newblock
\urldef\tempurl%
\url{https://pm.bsc.es/ftp/ompss-2/doc/spec/OmpSs-2-Specification.pdf}
\showURL{%
\tempurl}


\bibitem[\protect\citeauthoryear{BSC}{BSC}{2021}]%
        {bsc2019nanos6}
\bibfield{author}{\bibinfo{person}{BSC}.} \bibinfo{year}{2021}\natexlab{}.
\newblock \bibinfo{title}{Nanos6 Source}.
\newblock
\newblock
\urldef\tempurl%
\url{https://github.com/bsc-pm/nanos6}
\showURL{%
\tempurl}


\bibitem[\protect\citeauthoryear{Chamberlain}{Chamberlain}{2015}]%
        {chapel-coforall}
\bibfield{author}{\bibinfo{person}{Bradford~L Chamberlain}.}
  \bibinfo{year}{2015}\natexlab{}.
\newblock \showarticletitle{Chapel}.
\newblock In \bibinfo{booktitle}{\emph{Programming Models for Parallel
  Computing}}, \bibfield{editor}{\bibinfo{person}{Pavan Balaji}} (Ed.).
  \bibinfo{publisher}{MIT Press}, Chapter~6, \bibinfo{pages}{129--159}.
\newblock


\bibitem[\protect\citeauthoryear{Contreras and Martonosi}{Contreras and
  Martonosi}{2008}]%
        {tbb-oh}
\bibfield{author}{\bibinfo{person}{Gilberto Contreras} {and}
  \bibinfo{person}{Margaret Martonosi}.} \bibinfo{year}{2008}\natexlab{}.
\newblock \showarticletitle{Characterizing and improving the performance of
  Intel Threading Building Blocks}. In \bibinfo{booktitle}{\emph{2008 IEEE
  International Symposium on Workload Characterization}}.
  \bibinfo{pages}{57--66}.
\newblock
\urldef\tempurl%
\url{https://doi.org/10.1109/IISWC.2008.4636091}
\showDOI{\tempurl}


\bibitem[\protect\citeauthoryear{Corporation}{Corporation}{2022}]%
        {onetbb}
\bibfield{author}{\bibinfo{person}{Intel Corporation}.}
  \bibinfo{year}{2022}\natexlab{}.
\newblock \bibinfo{title}{oneAPI Threading Building Blocks (oneTBB)}.
\newblock
\newblock
\urldef\tempurl%
\url{https://oneapi-src.github.io/oneTBB/index.html}
\showURL{%
\tempurl}


\bibitem[\protect\citeauthoryear{Evans}{Evans}{2006}]%
        {jemalloc}
\bibfield{author}{\bibinfo{person}{Jason Evans}.}
  \bibinfo{year}{2006}\natexlab{}.
\newblock \showarticletitle{A Scalable Concurrent malloc(3) Implementation for
  FreeBSD}. In \bibinfo{booktitle}{\emph{BSDCan}} (Ottawa, Ontario, Canada).
\newblock
\urldef\tempurl%
\url{https://people.freebsd.org/~jasone/jemalloc/bsdcan2006/jemalloc.pdf}
\showURL{%
\tempurl}


\bibitem[\protect\citeauthoryear{Ferrer, Royuela, Caballero, Duran, Martorell,
  and Ayguad{\'e}}{Ferrer et~al\mbox{.}}{2011}]%
        {mercurium}
\bibfield{author}{\bibinfo{person}{Roger Ferrer}, \bibinfo{person}{Sara
  Royuela}, \bibinfo{person}{Diego Caballero}, \bibinfo{person}{Alejandro
  Duran}, \bibinfo{person}{Xavier Martorell}, {and} \bibinfo{person}{Eduard
  Ayguad{\'e}}.} \bibinfo{year}{2011}\natexlab{}.
\newblock \showarticletitle{Mercurium: Design decisions for a s2s compiler}. In
  \bibinfo{booktitle}{\emph{Cetus Users and Compiler Infastructure Workshop in
  conjunction with PACT}}.
\newblock


\bibitem[\protect\citeauthoryear{Frigo, Leiserson, and Randall}{Frigo
  et~al\mbox{.}}{1998}]%
        {cilk5}
\bibfield{author}{\bibinfo{person}{Matteo Frigo}, \bibinfo{person}{Charles~E.
  Leiserson}, {and} \bibinfo{person}{Keith~H. Randall}.}
  \bibinfo{year}{1998}\natexlab{}.
\newblock \showarticletitle{The Implementation of the Cilk-5 Multithreaded
  Language}. In \bibinfo{booktitle}{\emph{Proceedings of the ACM SIGPLAN 1998
  Conference on Programming Language Design and Implementation}} (Montreal,
  Quebec, Canada) \emph{(\bibinfo{series}{PLDI '98})}.
  \bibinfo{publisher}{Association for Computing Machinery},
  \bibinfo{address}{New York, NY, USA}, \bibinfo{pages}{212–223}.
\newblock
\showISBNx{0897919874}
\urldef\tempurl%
\url{https://doi.org/10.1145/277650.277725}
\showDOI{\tempurl}


\bibitem[\protect\citeauthoryear{Gautier, Perez, and Richard}{Gautier
  et~al\mbox{.}}{2018}]%
        {OpenMPGranularities}
\bibfield{author}{\bibinfo{person}{Thierry Gautier}, \bibinfo{person}{Christian
  Perez}, {and} \bibinfo{person}{J{\'e}r{\^o}me Richard}.}
  \bibinfo{year}{2018}\natexlab{}.
\newblock \showarticletitle{On the Impact of OpenMP Task Granularity}. In
  \bibinfo{booktitle}{\emph{Evolving OpenMP for Evolving Architectures}},
  \bibfield{editor}{\bibinfo{person}{Bronis~R. de~Supinski},
  \bibinfo{person}{Pedro Valero-Lara}, \bibinfo{person}{Xavier Martorell},
  \bibinfo{person}{Sergi Mateo~Bellido}, {and} \bibinfo{person}{Jesus Labarta}}
  (Eds.). \bibinfo{publisher}{Springer International Publishing},
  \bibinfo{address}{Cham}, \bibinfo{pages}{205--221}.
\newblock
\showISBNx{978-3-319-98521-3}


\bibitem[\protect\citeauthoryear{Heroux and Dongarra}{Heroux and
  Dongarra}{2013}]%
        {hpcg}
\bibfield{author}{\bibinfo{person}{Michael~Allen Heroux} {and}
  \bibinfo{person}{Jack. Dongarra}.} \bibinfo{year}{2013}\natexlab{}.
\newblock \showarticletitle{Toward a new metric for ranking high performance
  computing systems.}
\newblock  (\bibinfo{date}{6} \bibinfo{year}{2013}).
\newblock
\urldef\tempurl%
\url{https://doi.org/10.2172/1089988}
\showDOI{\tempurl}


\bibitem[\protect\citeauthoryear{Kruskal and Smith}{Kruskal and Smith}{1988}]%
        {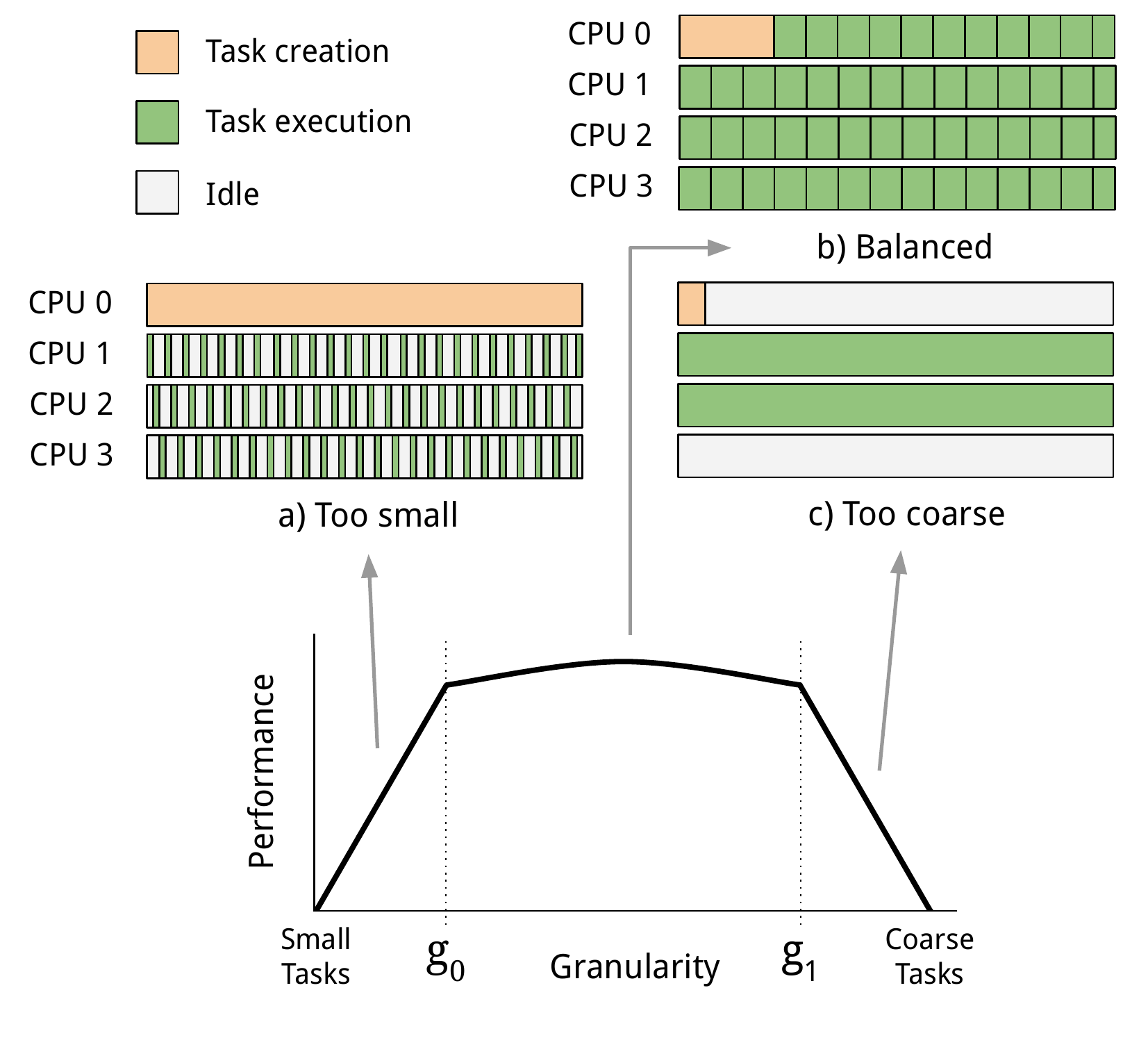}
\bibfield{author}{\bibinfo{person}{Clyde~P. Kruskal} {and}
  \bibinfo{person}{Carl~H. Smith}.} \bibinfo{year}{1988}\natexlab{}.
\newblock \showarticletitle{On the notion of granularity}.
\newblock \bibinfo{journal}{\emph{The Journal of Supercomputing}}
  \bibinfo{volume}{1}, \bibinfo{number}{4} (\bibinfo{date}{Aug.}
  \bibinfo{year}{1988}), \bibinfo{pages}{395--408}.
\newblock
\urldef\tempurl%
\url{https://doi.org/10.1007/bf00128489}
\showDOI{\tempurl}


\bibitem[\protect\citeauthoryear{{Maroñas}, {Sala}, {Mateo}, {Ayguadé}, and
  {Beltran}}{{Maroñas} et~al\mbox{.}}{2019}]%
        {WorksharingMarcos}
\bibfield{author}{\bibinfo{person}{M. {Maroñas}}, \bibinfo{person}{K. {Sala}},
  \bibinfo{person}{S. {Mateo}}, \bibinfo{person}{E. {Ayguadé}}, {and}
  \bibinfo{person}{V. {Beltran}}.} \bibinfo{year}{2019}\natexlab{}.
\newblock \showarticletitle{Worksharing Tasks: An Efficient Way to Exploit
  Irregular and Fine-Grained Loop Parallelism}. In
  \bibinfo{booktitle}{\emph{2019 IEEE 26th International Conference on High
  Performance Computing, Data, and Analytics (HiPC)}}.
  \bibinfo{pages}{383--394}.
\newblock
\showISSN{1094-7256}
\urldef\tempurl%
\url{https://doi.org/10.1109/HiPC.2019.00053}
\showDOI{\tempurl}


\bibitem[\protect\citeauthoryear{Maroñas, Teruel, and Beltran}{Maroñas
  et~al\mbox{.}}{2021}]%
        {taskloopdeps}
\bibfield{author}{\bibinfo{person}{Marcos Maroñas}, \bibinfo{person}{Xavier
  Teruel}, {and} \bibinfo{person}{Vicenç Beltran}.}
  \bibinfo{year}{2021}\natexlab{}.
\newblock \showarticletitle{OpenMP Taskloop Dependences}. In
  \bibinfo{booktitle}{\emph{OpenMP: Memory, Devices, and Tasks - 17th
  International Workshop on OpenMP, {IWOMP} 2021, Bristol, UK, October 14-16
  September, 2021, Proceedings}} \emph{(\bibinfo{series}{Lecture Notes in
  Computer Science})}, \bibfield{editor}{\bibinfo{person}{Naoya Maruyama},
  \bibinfo{person}{Bronis~R. de~Supinski}, {and} \bibinfo{person}{Mohamed
  Wahib}} (Eds.).
\newblock


\bibitem[\protect\citeauthoryear{Muddukrishna, Jonsson, and
  Brorsson}{Muddukrishna et~al\mbox{.}}{2016}]%
        {locality-sched}
\bibfield{author}{\bibinfo{person}{Ananya Muddukrishna},
  \bibinfo{person}{Peter~A. Jonsson}, {and} \bibinfo{person}{Mats Brorsson}.}
  \bibinfo{year}{2016}\natexlab{}.
\newblock \showarticletitle{Locality-Aware Task Scheduling and Data
  Distribution for OpenMP Programs on NUMA Systems and Manycore Processors}.
\newblock \bibinfo{journal}{\emph{Sci. Program.}}  \bibinfo{volume}{2015},
  Article \bibinfo{articleno}{5} (\bibinfo{date}{Jan.} \bibinfo{year}{2016}),
  \bibinfo{numpages}{1}~pages.
\newblock
\showISSN{1058-9244}
\urldef\tempurl%
\url{https://doi.org/10.1155/2015/981759}
\showDOI{\tempurl}


\bibitem[\protect\citeauthoryear{Navarro, Mateo, Perez, Beltran, and
  Ayguad{\'e}}{Navarro et~al\mbox{.}}{2017}]%
        {NavarroGranularity}
\bibfield{author}{\bibinfo{person}{Antoni Navarro}, \bibinfo{person}{Sergi
  Mateo}, \bibinfo{person}{Josep~Maria Perez}, \bibinfo{person}{Vicen{\c{c}}
  Beltran}, {and} \bibinfo{person}{Eduard Ayguad{\'e}}.}
  \bibinfo{year}{2017}\natexlab{}.
\newblock \showarticletitle{Adaptive and Architecture-Independent Task
  Granularity for Recursive Applications}. In \bibinfo{booktitle}{\emph{Scaling
  OpenMP for Exascale Performance and Portability}},
  \bibfield{editor}{\bibinfo{person}{Bronis~R. de~Supinski},
  \bibinfo{person}{Stephen~L. Olivier}, \bibinfo{person}{Christian Terboven},
  \bibinfo{person}{Barbara~M. Chapman}, {and} \bibinfo{person}{Matthias~S.
  M{\"u}ller}} (Eds.). \bibinfo{publisher}{Springer International Publishing},
  \bibinfo{address}{Cham}, \bibinfo{pages}{169--182}.
\newblock
\showISBNx{978-3-319-65578-9}


\bibitem[\protect\citeauthoryear{NVIDIA}{NVIDIA}{2021}]%
        {cuda}
\bibfield{author}{\bibinfo{person}{NVIDIA}.} \bibinfo{year}{2021}\natexlab{}.
\newblock \bibinfo{title}{CUDA C Programming Guide}.
\newblock
\newblock
\urldef\tempurl%
\url{https://docs.nvidia.com/cuda/cuda-c-programming-guide/}
\showURL{%
\tempurl}


\bibitem[\protect\citeauthoryear{{OpenMP Architecture Review Board}}{{OpenMP
  Architecture Review Board}}{2010}]%
        {openmp51}
\bibfield{author}{\bibinfo{person}{{OpenMP Architecture Review Board}}.}
  \bibinfo{year}{2010}\natexlab{}.
\newblock \bibinfo{title}{{OpenMP Technical Report 8: Version 5.1 Preview}}.
\newblock
\newblock
\urldef\tempurl%
\url{https://www.openmp.org/wp-content/uploads/openmp-TR8.pdf}
\showURL{%
\tempurl}
\newblock
\shownote{Accessed: 2020-02-01.}


\bibitem[\protect\citeauthoryear{Orozco, Garcia, Pavel, Khan, and Gao}{Orozco
  et~al\mbox{.}}{2013}]%
        {polytasks}
\bibfield{author}{\bibinfo{person}{Daniel Orozco}, \bibinfo{person}{Elkin
  Garcia}, \bibinfo{person}{Robert Pavel}, \bibinfo{person}{Rishi Khan}, {and}
  \bibinfo{person}{Guang~R. Gao}.} \bibinfo{year}{2013}\natexlab{}.
\newblock \showarticletitle{Polytasks: A Compressed Task Representation for HPC
  Runtimes}. In \bibinfo{booktitle}{\emph{Languages and Compilers for Parallel
  Computing}}, \bibfield{editor}{\bibinfo{person}{Sanjay Rajopadhye} {and}
  \bibinfo{person}{Michelle Mills~Strout}} (Eds.). \bibinfo{publisher}{Springer
  Berlin Heidelberg}, \bibinfo{address}{Berlin, Heidelberg},
  \bibinfo{pages}{268--282}.
\newblock
\showISBNx{978-3-642-36036-7}


\bibitem[\protect\citeauthoryear{Podobas, Brorsson, and Vlassov}{Podobas
  et~al\mbox{.}}{2014}]%
        {TurboBLYSK}
\bibfield{author}{\bibinfo{person}{Artur Podobas}, \bibinfo{person}{Mats
  Brorsson}, {and} \bibinfo{person}{Vladimir Vlassov}.}
  \bibinfo{year}{2014}\natexlab{}.
\newblock \showarticletitle{TurboB{\L}YSK: Scheduling for Improved Data-Driven
  Task Performance with Fast Dependency Resolution}. In
  \bibinfo{booktitle}{\emph{Using and Improving OpenMP for Devices, Tasks, and
  More}}, \bibfield{editor}{\bibinfo{person}{Luiz DeRose},
  \bibinfo{person}{Bronis~R. de~Supinski}, \bibinfo{person}{Stephen~L.
  Olivier}, \bibinfo{person}{Barbara~M. Chapman}, {and}
  \bibinfo{person}{Matthias~S. M{\"u}ller}} (Eds.).
  \bibinfo{publisher}{Springer International Publishing},
  \bibinfo{address}{Cham}, \bibinfo{pages}{45--57}.
\newblock
\showISBNx{978-3-319-11454-5}


\bibitem[\protect\citeauthoryear{Ros\`{a}, Rosales, and Binder}{Ros\`{a}
  et~al\mbox{.}}{2019}]%
        {GranularityAnalysis2019}
\bibfield{author}{\bibinfo{person}{Andrea Ros\`{a}}, \bibinfo{person}{Eduardo
  Rosales}, {and} \bibinfo{person}{Walter Binder}.}
  \bibinfo{year}{2019}\natexlab{}.
\newblock \showarticletitle{Analysis and Optimization of Task Granularity on
  the Java Virtual Machine}.
\newblock \bibinfo{journal}{\emph{ACM Trans. Program. Lang. Syst.}}
  \bibinfo{volume}{41}, \bibinfo{number}{3}, Article \bibinfo{articleno}{19}
  (\bibinfo{date}{July} \bibinfo{year}{2019}), \bibinfo{numpages}{47}~pages.
\newblock
\showISSN{0164-0925}
\urldef\tempurl%
\url{https://doi.org/10.1145/3338497}
\showDOI{\tempurl}


\bibitem[\protect\citeauthoryear{Soi, Bauer, Treichler, Papadakis, Lee,
  McCormick, Aiken, and Slaughter}{Soi et~al\mbox{.}}{2021}]%
        {indexlaunch}
\bibfield{author}{\bibinfo{person}{Rupanshu Soi}, \bibinfo{person}{Michael
  Bauer}, \bibinfo{person}{Sean Treichler}, \bibinfo{person}{Manolis
  Papadakis}, \bibinfo{person}{Wonchan Lee}, \bibinfo{person}{Patrick
  McCormick}, \bibinfo{person}{Alex Aiken}, {and} \bibinfo{person}{Elliott
  Slaughter}.} \bibinfo{year}{2021}\natexlab{}.
\newblock \showarticletitle{Index Launches: Scalable, Flexible Representation
  of Parallel Task Groups}. In \bibinfo{booktitle}{\emph{Proceedings of the
  International Conference for High Performance Computing, Networking, Storage
  and Analysis}} (St. Louis, Missouri) \emph{(\bibinfo{series}{SC '21})}.
  \bibinfo{publisher}{Association for Computing Machinery},
  \bibinfo{address}{New York, NY, USA}, Article \bibinfo{articleno}{66},
  \bibinfo{numpages}{18}~pages.
\newblock
\showISBNx{9781450384421}
\urldef\tempurl%
\url{https://doi.org/10.1145/3458817.3476175}
\showDOI{\tempurl}


\bibitem[\protect\citeauthoryear{Yang and He}{Yang and He}{2018}]%
        {Yang2018}
\bibfield{author}{\bibinfo{person}{Jixiang Yang} {and} \bibinfo{person}{Qingbi
  He}.} \bibinfo{year}{2018}\natexlab{}.
\newblock \showarticletitle{Scheduling Parallel Computations by Work Stealing:
  A Survey}.
\newblock \bibinfo{journal}{\emph{International Journal of Parallel
  Programming}} \bibinfo{volume}{46}, \bibinfo{number}{2} (\bibinfo{date}{01
  Apr} \bibinfo{year}{2018}), \bibinfo{pages}{173--197}.
\newblock
\showISSN{1573-7640}
\urldef\tempurl%
\url{https://doi.org/10.1007/s10766-016-0484-8}
\showDOI{\tempurl}


\bibitem[\protect\citeauthoryear{Yu, Royuela, and Qui{\~{n}}ones}{Yu
  et~al\mbox{.}}{2021}]%
        {taskgraph}
\bibfield{author}{\bibinfo{person}{Chenle Yu}, \bibinfo{person}{Sara Royuela},
  {and} \bibinfo{person}{Eduardo Qui{\~{n}}ones}.}
  \bibinfo{year}{2021}\natexlab{}.
\newblock \showarticletitle{Enhancing OpenMP Tasking Model: Performance and
  Portability}. In \bibinfo{booktitle}{\emph{OpenMP: Enabling Massive
  Node-Level Parallelism}}, \bibfield{editor}{\bibinfo{person}{Simon
  McIntosh-Smith}, \bibinfo{person}{Bronis~R. de~Supinski}, {and}
  \bibinfo{person}{Jannis Klinkenberg}} (Eds.). \bibinfo{publisher}{Springer
  International Publishing}, \bibinfo{address}{Cham}, \bibinfo{pages}{35--49}.
\newblock
\showISBNx{978-3-030-85262-7}


\end{thebibliography}

\end{document}